\documentclass[
reprint,
superscriptaddress,
amsmath,amssymb,
pra,
]{revtex4-2}
\usepackage{graphicx}
\usepackage{dcolumn}
\usepackage{bm}
\usepackage{hyperref}
\usepackage[dvipsnames]{xcolor}
\usepackage{soul}
\usepackage{ulem}
\usepackage[left]{lineno}
\usepackage{braket}
\usepackage{mathdots}
\usepackage{MnSymbol}

\definecolor{dark-green}{RGB}{0, 128, 0}

\begin{document}

\title{Decoherence-free subspaces in the noisy dynamics of discrete-step quantum walks\\
in a photonic lattice}

\author{Rajesh Asapanna}
\affiliation{Univ. Lille, CNRS, UMR 8523 -- PhLAM -- Physique des Lasers Atomes et Mol\'ecules, F-59000 Lille, France}

\author{Clément~Hainaut}
\affiliation{Univ. Lille, CNRS, UMR 8523 -- PhLAM -- Physique des Lasers Atomes et Mol\'ecules, F-59000 Lille, France}

\author{Alberto~Amo}
\email{alberto.amo-garcia@univ-lille.fr}
\affiliation{Univ. Lille, CNRS, UMR 8523 -- PhLAM -- Physique des Lasers Atomes et Mol\'ecules, F-59000 Lille, France}

\author{\'Alvaro~G\'omez-Le\'on}
\email{a.gomez.leon@csic.es}
\affiliation{Institute of Fundamental Physics IFF-CSIC, Calle Serrano 113b, 28006 Madrid, Spain}

\date{\today}

\begin{abstract}
We study the noisy dynamics of periodically driven, discrete-step quantum walks in a one-dimensional photonic lattice. We find that in the bulk, temporal noise that is constant within a Floquet period leads to decoherence-free momentum subspaces, whereas fully random noise destroys coherence in a few time-steps. When considering topological edge states, we observe decoherence no matter the type of temporal noise. To explain these results, we derive a non-perturbative master equation to describe the system’s dynamics. We confirm experimentally our findings in a time multiplexed photonic lattice implemented in a double-fibre ring setup subject to laser pulse input states, in which we engineer different types of temporal noise.
\end{abstract}
\maketitle
Quantum walks~\cite{aharonov_quantum_1993}, the quantum analog of classical random walks, have emerged as a powerful framework for exploring quantum transport~\cite{Mulken2011,Schreiber2011}, designing quantum algorithms~\cite{Ambainis2003,shenvi_quantum_2003,childs_universal_2009}, and simulating complex quantum systems~\cite{Stamp_QWalks_Decoherence_2006,Stamp_QWalks_Encoding_2007,Schreiber_2D_QW_2012,lee_quantum_2015,vakulchyk_wave_2019}. 
Among these, discrete-step quantum walks (DSQWs) in which an initial state evolves in a lattice of waveguides subject to a cascade of discrete unitary operations, have proven highly adaptable for both theoretical investigations and experimental realizations. They provide a way to coherently manipulate quantum states over many time steps~\cite{kok_linear_2007,nahum_operator_2018,monika_quantum_2025}, and they are perfectly suited for simulating condensed-matter phenomena such as ballistic spreading~\cite{schreiber_photons_2010}, Bloch oscillations~\cite{Wimmer2015, adiyatullin_topological_2022, hu_observing_2024, zhao_blochzener_2024}, localization~\cite{Schreiber2011}, gauge fields~\cite{Chalabi2019, ye_reconfigurable_2023, lin_manipulating_2023}, topological phases~\cite{kitagawa_observation_2012, Bisianov2019} and non-Hermitian dynamics~\cite{mochizuki_explicit_2016, xiao_higher_2018, xiao_non-hermitian_2020,Weidemann2020, mittal_persistence_2021}.

The discrete-step nature of DSQWs renders them ideally suited for modelling Floquet lattices subject to periodic driving, which have been shown to lead to exotic nonequilibrium phases without a static equivalent, such as anomalous Floquet topological insulators~\cite{kitagawa_observation_2012, Cardano2017, Maczewsky2017, Mukherjee2017a, cheng_observation_2019}.
Furthermore, the step-by-step evolution in DSQWs introduces additional topological features such as winding bands\cite{adiyatullin_topological_2022} and extrinsic topology~\cite{bessho_extrinsic_2022, el_sokhen_edge-dependent_2024,asapanna_observation_2024}, which allows controlling the number of topological edge states without requiring changes in the bulk.

However, real-world quantum systems are inevitably exposed to environmental noise and decoherence. Understanding how noise affects DSQWs is essential both for their technological deployment and for establishing their robustness as quantum simulators. Noise can degrade coherence, suppress interference, delocalize edge modes, and induce classical dynamics, thereby limiting the quantum walk's ability to outperform classical counterparts~\cite{Shapira2003,Chandrashekar2007,Yin2008,rieder_localization_2018,sieberer_statistical_2018}. Consequently, studying noisy DSQWs is not only crucial for practical quantum technologies but also serves as a theoretical laboratory for exploring the interplay between decoherence, periodic driving, and topological protection.

In this work, we investigate the dynamics of DSQWs in a one-dimensional lattice in the presence of temporal noise, focusing on how the dynamics of initially localized wavepackets is affected both in the bulk and in topological edge states. 
We consider a lattice of cascaded unitary operators in which DSQWs are periodic in time with a characteristic Floquet period.
We derive a non-perturbative expression for the master equation governing the dynamics of the density matrix, averaged over noise realizations. 
This allows us to show that it is possible to find controllable decoherence-free subspaces in momentum space for the bulk dynamics when the noise is constant within a Floquet period. 
If noise is random also within the Floquet period, it destroys coherence in a few time steps. 
In the case of a topological edge state, we observe decoherence regardless of the type of temporal noise. 

We experimentally demonstrate our findings by realizing a noisy DSQWs for light pulses in a setup of two coupled fiber loops.~\cite{Regensburger2011, Bisianov2019, el_sokhen_edge-dependent_2024}. 
This time-multiplexed architecture allows for large-scale quantum walks in a lattice with high stability and excellent control over both unitary operations and engineered noise. Even though our experiments are done in the classical regime using coherent laser sources as input states, the results can be applied to quantum communications and photonic quantum information processing~\cite{monika_quantum_2025}.

The manuscript is structured as follows. Section~\ref{sec:bulk} describes the discrete-step lattice model and the effect of different types of noise in the bulk of the lattice. Section~\ref{sec:edge} addresses the decoherence of topological edge states in presence of noise. The final section discusses implications of this work.

\begin{figure}[t!]
\includegraphics[width=\columnwidth]{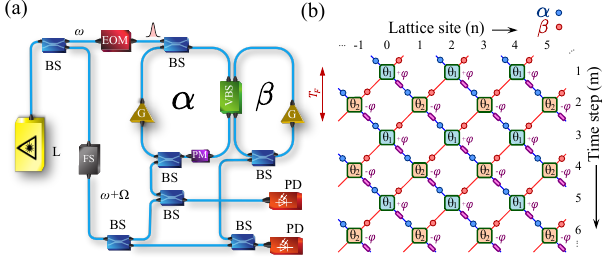}
\caption{\label{fig:expt_setup} 
(a) Scheme of the experimental setup with beam splitters BS, variable beam splitter VBS, electro-optic modulator EOM, phase modulator PM, photo-diodes PD, amplifier G and frequency shifter FS to create a local oscillator for the measurement of the eigenvectors and eigenvalues. The $\alpha$ and $\beta$ rings have a length of 45.34 m and 44.63 m, respectively. 
(b)~Discrete-step lattice after time-demultiplexing of the pulses in the double ring with $\theta$ and $\varphi$ correspond to couplings from VBS and phase from PM, respectively.
}
\end{figure}
\section{Bulk noisy dynamics}\label{sec:bulk}

To investigate the effect of temporal noise in DSQWs, we consider a lattice of cascaded unitary operators like the one illustrated in Fig.~\ref{fig:expt_setup}(b).
It can be experimentally implemented in the setup of Fig.~\ref{fig:expt_setup}(a). It consists of two fiber loops of slightly different lengths, coupled through a variable beamsplitter.
A short square laser pulse of approximately 1.4~ns at a wavelength of 1550~nm is injected into the $\alpha$ loop, where it undergoes split-step walk dynamics each time it encounters the beamsplitter.
The slight length difference between the loops encodes the lattice position $n$ in the pulse arrival time at the output port during each round trip~\cite{Regensburger2011}. 
The round-trip duration defines the discrete time-step index $m$. 
Erbium-doped fiber amplifiers integrated into the loops compensate for insertion, extraction, and propagation losses, enabling pulses to circulate over many round trips.
In addition, a phase modulator introduced in the $\alpha$ ring allows adding an extra controlled phase which can be used to engineer the lattice dispersion.

The dynamics in the fibers maps directly onto the discrete-step evolution of light pulses in the one-dimensional lattice displayed in Fig.~\ref{fig:expt_setup}(b). 
It represents a generic one-dimensional lattice model in which an input state distributed over various initial sites $n$ evolves in discrete steps in time $m$. 
At each step, the state follows a unitary operation of splitting and phase shift (at the phase modulator PM shown in the $\alpha$ ring in Fig.~\ref{fig:expt_setup}(a)). 
This kind of evolution is typical of a cascade of beam splitters in optics, and has been implemented using fiber ring setups similar to ours to study different aspects of lattice dynamics~\cite{Schreiber2011, Regensburger2011, weidemann_topological_2020, Chalabi2019, adiyatullin_topological_2022, ye_reconfigurable_2023, lin_manipulating_2023, dinani_universal_2025}.
It has also been implemented in coin polarization walkers~\cite{kitagawa_observation_2012, wang_detecting_2018, xu_measuring_2018}, integrated photonic circuits~\cite{bogaerts_programmable_2020}, and coupled photonic resonators~\cite{afzal_realization_2020}. 
Most of these publications report the dynamics of laser pulses with a large number of photons, which can be described in the classical regime.
A few works have used the same kind of discrete-step lattices in the quantum regime of few photons to investigate quantum dynamics of single photons and entangled states~\cite{Schreiber2011, wang_detecting_2018, monika_quantum_2025}.

The time evolution of light pulses follows the set of coupled equations~\cite{Regensburger2011}:
\begin{align}
\alpha_{n}^{m+1} & =\left[\alpha_{n-1}^{m}\cos\left(\theta_{m}\right)+i\beta_{n-1}^{m}\sin\left(\theta_{m}\right)\right]e^{i\varphi_{m}},\nonumber\\
\beta_{n}^{m+1} & =i\alpha_{n+1}^{m}\sin\left(\theta_{m}\right)+\beta_{n+1}^{m}\cos\left(\theta_{m}\right),\label{eq:BasicEq}
\end{align}
where $\alpha_{n}^{m}$ and $\beta_{n}^{m}$ describe the complex amplitude of the pulses in the long and short rings, respectively, at time step $m$ and lattice site $n$. 
The coupling angles $\theta_{m}$ and the phases $\varphi_{m}$ introduced by a phase modulator in the $\alpha$ ring can be dynamically modulated electronically.
The modulated phase $\varphi_{m}$ alternates between $+\varphi$ and $-\varphi$ at odd and even time steps. Its value allows engineering the band structure of the lattice with flexibility~\cite{Wimmer2017, ye_reconfigurable_2023}. When discussing the theoretical model below, we keep the general form of the lattice model with $\varphi_{m}$. In most of the experiments, its value will be equal to zero; only in the final part of this work, when considering flat band models with topological edge states, we will implement a lattice with $\varphi_{m}\neq 0$.

The evolution of an initial state $|\psi_0\rangle$ at time step $m$ is expressed as $|\psi_m\rangle= \hat{U}_m\ldots \hat{U}_0|\psi_0\rangle$, where $\hat{U}_j$ denotes the evolution operator at time $j$. 
Because each lattice site has two components, the $\alpha$ and $\beta$ sublattice sites, the evolution operator $\hat{U}_j$ is a $2N\times 2N$ matrix. However, for the bulk dynamics one can perform a Fourier transform of Eq.~\eqref{eq:BasicEq} to momentum representation and reduce $\hat{U}_j$ to a $2 \times 2$ matrix with its explicit form given in Appendix~\ref{sec:Appendix1}.
For periodic protocols (i.e., when $\theta_m$ varies periodically), one can describe the stroboscopic dynamics at times corresponding to the driving period by using the Floquet operator $\hat{U}_{F}=\hat{U}_T\ldots \hat{U}_1$, where $T$ is the number of steps in a period of the driving protocol.

We now artificially introduce uncorrelated noise in the variable beamsplitter at different time steps $m$, in the form $\theta_{m} \to \theta_{m}+\tau_{m}$, with $\tau_m$ being random numbers from a normal distribution with zero mean, $\overline{\tau}_{m}=0$ and no correlation $\overline{\tau_{m}\tau_{m^{\prime}}} = \sigma^{2}\delta_{m,m^{\prime}}$. 
This kind of noise generally breaks the discrete time-translation symmetry of the protocol and makes Floquet theorem to break down.

To study the noisy dynamics of Eq.~\eqref{eq:BasicEq} we derive a master equation that describes the system's density matrix, averaged over noise realizations.
A related analysis was done in Ref.~\cite{sieberer_statistical_2018} for random noise and to second order perturbation theory, leading to a Lindblad form of the master equation.
Instead, here we perform a full non-perturbative analysis by re-summation of all orders of the perturbative series. We find that despite the master equation not being in obvious Lindblad form, it correctly predicts the dynamics and incorporates additional higher noise-induced processes that are missing in the perturbative description. This is important to ensure the existence of decoherence free subspaces below.
\subsection*{Stroboscopic noise}
We first consider the specific situation of noise that modifies in the same way all splitting angles $\theta_m$ within a Floquet period $M$: within period $M$, all $\theta_m$ take the form $\theta_m+\tau_M$. We call this situation \textit{stroboscopic noise}. We study the density matrix that accounts for the dynamics at each period $M$, $\hat{\rho}_M=|\psi_{M}\rangle\langle\psi_{M}|$, which is connected to the previous period :
\begin{equation}\label{eq:desnsity_strob}
\hat{\rho}_{M+1} = \hat{U}_{F}\left(\tau_{M+1}\right)\hat{\rho}_M\hat{U}_{F}^{\dagger}\left(\tau_{M+1}\right),
\end{equation}
where $|\psi_{M}\rangle=\hat{U}_F(\tau_M) \ldots \hat{U}_F(\tau_2) \hat{U}_F(\tau_1)|\psi_{0}\rangle$ is the state of the system after $M$ stroboscopic periods and $\hat{U}_{F}\left(\tau_{M}\right)$ is the operator at period $M$.
Note that the density matrix $\hat{\rho}_M$ is a function of all previous noise values $\tau_{1},\tau_{2},\ldots,\tau_{M}$.
Each noise realization leads to unitary dynamics with a different trajectory. 
However, as we are interested in robust average properties, we calculate the averaged density matrix over noise realizations
$\overline{\hat{\rho}_M}$ by averaging Eq.~\eqref{eq:desnsity_strob}. 
Importantly, for uncorrelated noise we can factorize the noise averages at different time steps and this highly simplifies the calculation of the expectation value.

To calculate the noise average at a single time step, it is always possible to write the Floquet operator as a sum of matrices weighted by noise-dependent prefactors by doing a Taylor series expansion of $\hat{U}_{F}$ in $\tau_M$ and regrouping terms with identical powers:
\begin{equation}
    \hat{U}_{F}\left(\tau_{M}\right) = \sum_{\mu}f_{\mu}\left(\tau_{M}\right)\hat{\mathcal{U}}_{\mu}.\label{eq:UF-Decomposition}
\end{equation}
\noindent The index $\mu$ runs up to the cut-off of the Taylor expansion. 
In this formulation, the functions $f_{\mu}\left(\tau_{M}\right)$ contain the dependence on the noise variable $\tau_M$, while the matrices $\hat{\mathcal{U}}_{\mu}$ remain independent of it.
Some explicit examples are shown below and in Appendix~\ref{sec:Appendix1}.
Finally, in this form, it is possible to perform the average over noise realizations:
\begin{equation}
    \overline{\hat{\rho}_{M+1}} = \sum_{\mu,\nu}F_{\mu,\nu}\left(\sigma^{2}\right)\hat{\mathcal{U}}_{\mu}\overline{\hat{\rho}_M}\hat{\mathcal{U}}_{\nu}^{\dagger}\label{eq:MasterEq1}
\end{equation}
with $F_{\mu,\nu}\left(\sigma^{2}\right)=\overline{f_{\mu}\left(\tau_{M}\right)f_{\nu}\left(\tau_{M}\right)}$. Equation~\eqref{eq:MasterEq1} corresponds to the general form of the master equation for the noisy dynamics.  

In the following, we consider a two-step Floquet protocol of the form $\hat{U}_{F} = \hat{U}_{2}\hat{U}_{1}$, defined by splitting angles $\theta_{1}$ and $\theta_{2}$, and values of the phase modulator alternating at odd and even steps between $+\varphi$ and $-\varphi$.
If we first focus on the bulk, the quasienergy spectrum of the lattice without noise can be computed from the eigenvalues of the Floquet operator in momentum space $\hat{U}_{F} (k)$, which take the form (see Appendix~\ref{sec:AppendixEigen} for further details):
\begin{equation}\label{eq:eigenvalues}
E_{\pm}=\pm \arccos[\cos{\theta_1}\cos{\theta_2}\cos k-\sin{\theta_1}\sin{\theta_2}\cos\varphi].
\end{equation}
Due to Floquet periodicity, the eigenvalues exhibit two bands and two inequivalent gaps at energies $E=0$ and $E=\pi$.


To study of the effect of stroboscopic noise in the bulk dynamics, the Floquet operators become noise dependent and take the following form $\hat{U}_F (\tau_M) = \hat{U}_2(\tau_M) \hat{U}_1(\tau_M)$ at each Floquet period $M$.
The decomposition of the noisy Floquet operator in Eq.~(\ref{eq:UF-Decomposition}) can be exactly summed into in three different noise terms for a given momentum $k$:
\begin{equation}
    \hat{U}_{F}\left(k, \tau_{M}\right) = \hat{U}_{F}\left(k, 0\right)+f_{+}\left(\tau_{M}\right)\hat{\mathcal{U}}_{+}\left(k\right)+f_{-}\left(\tau_{M}\right)\hat{\mathcal{U}}_{-}\left(k\right),
\end{equation}
with $f_{-}\left(\tau_{M}\right) = -\sin\left(\tau_{M}\right) \cos\left(\tau_{M}\right)$ and $f_{+}\left(\tau_{M}\right) = -\sin^{2}\left(\tau_{M}\right)$. The matrices $\hat{\mathcal{U}}_{\pm}\left(k\right)$ are explicitly written in Appendix~\ref{sec:Appendix1}.
Performing the average over noise configurations we arrive at:
\begin{align}\label{eq:masterStrob}
    \overline{\hat{\rho}_{M+1}} =& \hat{U}_{F}\overline{\hat{\rho}_{M}}\hat{U}_{F}^{\dagger}  -\Gamma_{+}\left(\hat{\mathcal{U}}_{+}\overline{\hat{\rho}_{M}}\hat{U}_{F}^{\dagger}+\hat{U}_{F}\overline{\hat{\rho}_{M}}\hat{\mathcal{U}}_{+}^{\dagger}\right) \nonumber \\
    &+\Gamma_{+,+}\hat{\mathcal{U}}_{+}\overline{\hat{\rho}_{M}}\hat{\mathcal{U}}_{+}^{\dagger}    +\Gamma_{-,-}\hat{\mathcal{U}}_{-}\overline{\hat{\rho}_{M}}\hat{\mathcal{U}}_{-}^{\dagger},
\end{align}
where we have omitted the $k$ dependence, $\hat{U}_{F}$ is the noiseless Floquet operator, $\Gamma_{+} = (1-e^{-2\sigma^{2}})/2$, $\Gamma_{+,+} =(3+e^{-8\sigma^{2}}-4e^{-2\sigma^{2}})/8$ and $\Gamma_{-,-} = (1-e^{-8\sigma^{2}})/8$.
The first term describes the free evolution under a noiseless Floquet protocol, while the other terms characterize different decoherence processes due to noise fluctuations. 
All the noise dependence in Eq.~(\ref{eq:masterStrob}) is encoded in the $\Gamma$ prefactors (the matrices $\hat{U}_F$ and $\hat{\mathcal{U}}_\pm$ depend only on $\theta_{1,2}$, $\varphi$ and $k$), resulting in a decay of the coherence. 
Interestingly, a series expansion of $\Gamma_{+,+} \sim \sigma^4+\mathcal{O}(\sigma^6)$ shows that the lowest order contribution from this term is $\sigma^4$, indicating that perturbative expressions to second order, such as those used to obtain a Lindblad form, would miss the contribution from this noise process to the noise-averaged density matrix.

More importantly, the matrices $\hat{\mathcal{U}}_{\pm}$ are proportional to $\left(e^{\pm i k}+e^{\pm i \varphi}\right)$, which means that if $k=\varphi+(2p+1)\pi$ with $p \in \mathbb{Z}$, they vanish and Eq.~\eqref{eq:masterStrob} reduces to its noisless form. This means that a wavepacket with this value of momentum will not be affected by decoherence: in presence of stroboscopic noise, there are bulk states immune to temporal noise fluctuations. 
This can also be seen from the form of the spectrum in Eq.~\eqref{eq:eigenvalues}: for those values of $k$, it simplifies to $E_{\pm}=\pm \arccos[\cos\varphi\cos({\theta_1 - \theta_2})]$, and equal values of the noise at the two steps of each period are cancelled.
On the other hand, the matrices $\hat{\mathcal{U}}_{\pm}\sim \left(e^{\pm i k}+e^{\pm i \varphi}\right)$ have a maximal effect in inducing decoherence for values of $k=\varphi+2p\pi$.
Again, this can be understood from the form of the noiseless eigenvalues, which for those values of $k$ take the form $E_{\pm}=\pm \arccos[\cos\varphi\cos({\theta_1 + \theta_2})]$, highlighting that the stroboscopic noise affects maximally this momentum subspace.

Note that the decoherence-free subspaces we have just identified exist even if the splitting angle is fully random from period to period.
Such a situation could be used to manipulate and filter out information, which might be useful in certain applications.

\subsection*{Random noise}
This surprising result is a consequence of the stroboscopic nature of the noise being considered. 
To see this, we now consider the more general case in which noise $\tau_{m}$ in the splitting angle changes randomly at each step $m$.
In this case, the time evolution operator at each time step can be expressed as:
\begin{equation}
    \hat{U}_{m}\left(\tau_{m}\right) = \cos\left(\tau_{m}\right)\hat{U}_{m}+\sin\left(\tau_{m}\right)\hat{U}_{m}^\prime,
\end{equation}
where $\hat{U}_{m}$ is the noiseless step operator characterized by the noiseless splitting angles $\theta_1$ or $\theta_2$ at either odd or even steps $m$, and $\hat{U}_{m}^\prime = \left.\partial_{\tau_m}\hat{U}_{m}\left(\tau_m\right)\right|_{\tau_m=0}$. The master equation for one time step, obtained after noise averaging is:
\begin{equation}
    \overline{\hat{\rho}_{m+1}} = \frac{1+e^{-2\sigma^{2}}}{2}\hat{U}_{1}\overline{\hat{\rho}_{m}}\hat{U}_{1}^{\dagger} + \Gamma_+ \hat{U}_{1}^{\prime}\overline{\hat{\rho}_{m}}\hat{U}_{1}^{\prime\dagger}.
\end{equation}
Noise averaging now acts at each time step instead of at each stroboscopic time. 
For proper comparison with Eq.~(\ref{eq:masterStrob}) we look at the master equation after two steps i.e. single Floquet step of the protocol:
\begin{align}\label{eq:masterRand}
    \overline{\hat{\rho}_{m+2}} =&\left(\frac{1+e^{-2\sigma^{2}}}{2}\right)^{2}\hat{U}_{F}\overline{\hat{\rho}_{m}}\hat{U}_{F}^{\dagger}+\Gamma_{+}^{2}\hat{U}_{2}^{\prime}\hat{U}_{1}^{\prime}\overline{\hat{\rho}_{m}}\hat{U}_{1}^{\prime\dagger}\hat{U}_{2}^{\prime\dagger} \nonumber \\
    &+ \frac{1-e^{-4\sigma^{2}}}{4}\left(\hat{U}_{2}\hat{U}_{1}^{\prime}\overline{\hat{\rho}_{m}}\hat{U}_{1}^{\prime\dagger}\hat{U}_{2}^{\dagger}+\hat{U}_{2}^{\prime}\hat{U}_{1}\overline{\hat{\rho}_{m}}\hat{U}_{1}^{\dagger}\hat{U}_{2}^{\prime\dagger}\right).
\end{align}
Different from Eq.~(\ref{eq:masterStrob}), for $\sigma\neq 0$, the contribution of the first term to the dynamics is always modified by the noise. This means that decoherence-free subspaces do not exist for random noise.

\subsection*{Experimental results}
\begin{figure}[t!]
\includegraphics[width=\columnwidth]{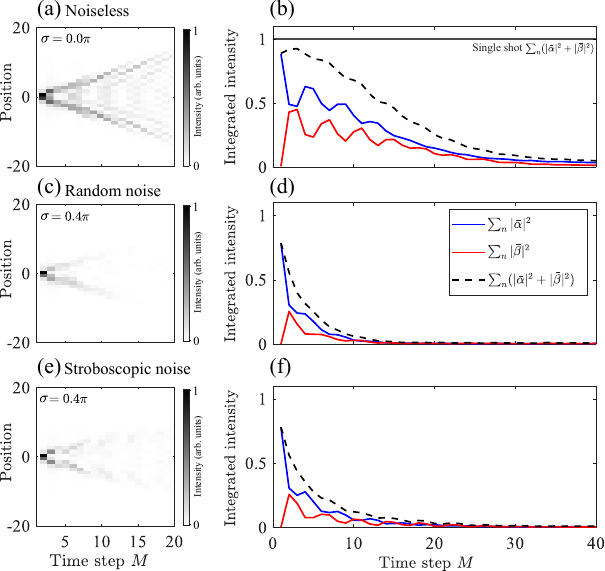}
\caption{\label{fig:spatio_temp} 
(a), (c), (e) Measured light intensity in the $\beta$ ring after single-site injection under different types of step noise on a discrete-step lattice with $\theta_1=0$, $\theta_2=0.25\pi$, and $\varphi=0$. 
The spatiotemporal dynamics have been coherently averaged ($|\mathrm{mean}(\beta)|^2$) over 100 noise realisations.
(a) Noiseless evolution ($\sigma=0$). 
(c) Evolution under large random noise ($\sigma=0.4\pi$). 
(e) Evolution under large stroboscopic noise ($\sigma=0.4\pi$). 
(b), (d), (f) Light intensity summed over all lattice sites from the measured averaged spatiotemporal dynamics in the $\alpha$ and $\beta$ rings in the conditions of (a), (c), (e).}
\end{figure}

These predictions are confirmed experimentally. Figures~\ref{fig:spatio_temp}(a), (c) and (d) display the measured spatiotemporal dynamics after injecting a pulse at a single site in the $\alpha$ ring at step $m=1$ for a lattice with $\theta_1=0$, $\theta_2=0.25\pi$, and $\varphi=0$.
The plots show the intensity $|\bar\beta_n^M|^2$ at each lattice site $n$ and Floquet step $M$ computed after averaging the measured evolution of the amplitude over 100 realizations. 
To do the coherent averaging over the 100 realizations, we need to measure the amplitude and phase at each site and time step for each individual realization.
The procedure to do this and to compute the band structure is described in detail in Appendix~\ref{Appendix:data_proc}.
The average measured intensity corresponds to the diagonal terms of the density matrix, which follow Eqs.~(\ref{eq:masterStrob}) and (\ref{eq:masterRand}).

In the absence of noise [Figure~\ref{fig:spatio_temp}(a)], spatiotemporal interference fringes characteristic of coherent evolution are observed during propagation.
They are more clearly seen in Fig.~\ref{fig:spatio_temp}(b) which shows the sum over all lattice sites of the intensity at each Floquet time step $M$ of the $\alpha$ (blue) and $\beta$ (red) sublattices. 
The dashed black line is the sum of the intensity on both sublattices $\sum_n |\bar\alpha_n^M|^2 + |\bar\beta_n^M|^2$, which should be equal to 1 at all time steps in the noiseless case. 
The reason for the observed decay of the averaged amplitudes is that the amplitude and phase extracted from each realization present experimental errors coming mostly from fiber length variations, electronic noise and data processing. 
When doing the coherent average over 100 realizations, these errors sum up to display the decays observed in Fig.~\ref{fig:spatio_temp}(b). 
The gray line shows the total intensity in the $\alpha$ and $\beta$ rings in a single realization, with no apparent decay.

Figure~\ref{fig:spatio_temp}(c)-(d) shows that random noise in the splitting angles fully destroys the spatiotemporal interference fringes, indicating noise-induced decoherence.
The decay of the averaged signal in time is also a signature of the decoherence and of the decay of the diagonal terms of the density matrix: the complex evolution of the amplitudes $\alpha_n^M$ and $\beta_n^M$ average out over many realizations of noise.
When the noise is stroboscopic, Fig.~\ref{fig:spatio_temp}(e)-(f) shows a similar decay of the coherent average but the interference patterns persist longer (Fig.~\ref{fig:spatio_temp}(f)) due to the presence of decoherence-free subspaces that protect certain momentum components.

\begin{figure}[t!]
\includegraphics[width=\columnwidth]{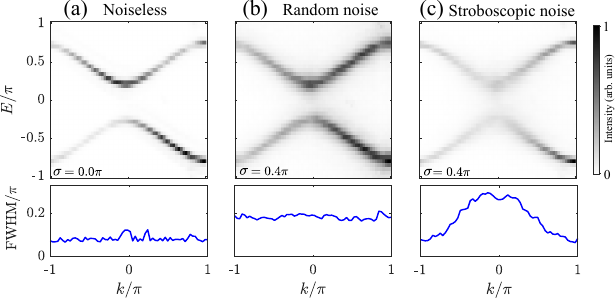}
\caption{\label{fig:bands} 
Measured dispersions under different types of noise for a lattice with $\theta_1=0$, $\theta_2=0.25\pi$, and $\varphi=0$. The intensity is computed by $|\tilde{\alpha}|^2 + |\tilde{\beta}|^2$, where the tilde indicates the Fourier amplitudes of the $\alpha$ and $\beta$ sublattices, and then it is averaged over 100 independent noise realizations.
The lower panel shows the Gaussian-fitted full width at half maximum (FWHM) of the upper band for each quasimomentum $k$ .
(a) Noiseless evolution ($\sigma=0$).
(b) Large random noise ($\sigma=0.4\pi$).
(c) Large stroboscopic noise ($\sigma=0.4\pi$).
}
\end{figure}

The decoherence-free subspaces can be clearly identified in momentum space.
Figure~\ref{fig:bands} shows the dispersion relation averaged over 100 realizations obtained via the Fourier transform of each individual measured spatiotemporal evolution for the lattice parameters of Fig.~\ref{fig:spatio_temp}. %
The noiseless case (Fig.~\ref{fig:bands}(a)) displays uniform broadening of the bands across $k$ (see lower panel), with a value mostly limited by the total number of steps in the experiment (80 steps).
Numerical simulations of Eq.~\ref{eq:BasicEq} in the conditions of Fig.~\ref{fig:bands}(a) show a full width at half maximum 70\% narrower than the experimental value; the discrepancy with the experiment must arise from experimental noise.
Introducing artificial random noise (Fig.~\ref{fig:bands}(b)) produces uniform broadening across all momenta.
In contrast, stroboscopic noise (Fig.~\ref{fig:bands}(c)) results in inhomogeneous broadening, with minimal broadening near $k = \pm \pi$, comparable to the noiseless case.
At $k = 0$ we observe an enlarged broadening as compared to the random noise case, as predicted by the analysis of the density matrix for the value of $k=\varphi +2p\pi=0$ (in our experiment $\varphi=0$, and we take $p=0$).

Note that in the experiment, the random variables $\tau_m$ are sampled from an uncorrelated uniform distribution $\tau_m \in [-\sigma/2, +\sigma/2]$, rather than a normal distribution (which was the case in the above theoretical computation of the density matrix). 
The decoherence-free subspaces emerge independently of the noise distribution, highlighting their robustness against different noise models.

\section{Edge dynamics}\label{sec:edge}
We now study how stroboscopic and non-stroboscopic noise affect topological edge states. These edge states appear in one-dimensional lattices for particular values of $\theta_1$ and $\theta_2$~\cite{Bisianov2019}.
Using a time-continuous Hamiltonian description, Ref.~\cite{rieder_localization_2018}  shows that random fluctuations result in an exponential decay of the edge state population. This decay could be transformed into a power law by introducing spatial localization of the bulk via the presence of flat bands or spatial disorder.

To study the dynamics of topological edge states under stroboscopic noise, we use the non-perturbative master equation, Eq.~\eqref{eq:MasterEq1}, but now in real space to account for the finite size lattice with $N$ sites.
Following a procedure analogous to the one leading to Eq.~(\ref{eq:UF-Decomposition}), the noisy Floquet operator in real space can be decomposed into a sum of five terms:
\begin{align}
    \hat{U}_{F}\left(\tau_{M}\right)=& \hat{\mathcal{U}}_{0}+\cos\left(\tau_{M}\right)\hat{\mathcal{U}}_{c}+ \sin\left(\tau_{M}\right)\hat{\mathcal{U}}_{s}+ \cos^{2}\left(\tau_{M}\right)\hat{\mathcal{U}}_{cc} \nonumber \\
    &+\sin\left(\tau_{M}\right)\cos\left(\tau_{M}\right)\hat{\mathcal{U}}_{sc}+ \sin^{2}\left(\tau_{M}\right)\hat{\mathcal{U}}_{ss}.
\end{align}
The method to compute matrices $\hat{\mathcal{U}}$ and the final form of the master equation after noise averaging can be found in Appendix~\ref{Appendix:master_edge}. 

\begin{figure}
    \centering
    \includegraphics[width=\columnwidth]{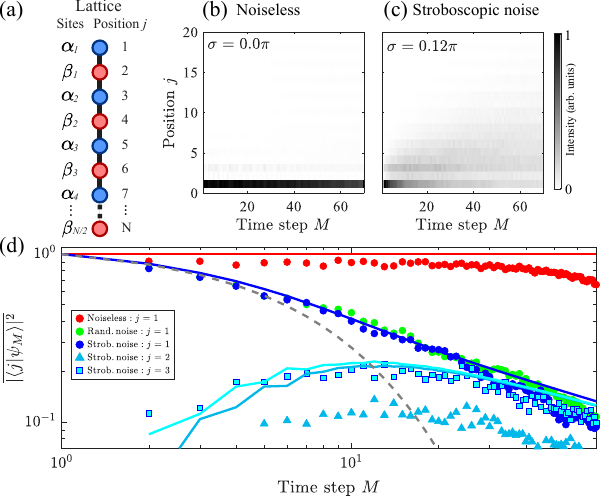}
    \caption{(a) Scheme of the labeling of the sites in a finite size lattice. The edge site is located at $j=1$. 
    (b) Measured spatiotemporal dynamics in a finite lattice with parameters $\theta_1=0.5\pi$, $\theta_2=0.0\pi$, $\varphi=0.2\pi$ and $N=44$ number of sites without added noise.
    The initial condition is excitation at the $j=1$ site.
    (c) Same as (b) in the presence of stroboscopic noise with $\sigma=0.12\pi$, averaged over 100 realizations following the procedure of Fig.~\ref{fig:spatio_temp}.
    (d) Measured (dots) averaged occupation probability $\overline{|\langle j | \psi_M \rangle|^2} $ as a function of step $M$ after initial excitation of
the left edge site $j=1$, under different types of noise with $\sigma=0.12\pi$.
    Blue triangles and squares show the evolution of bulk sites.
    For comparison, the gray dashed line indicates purely exponential decay with rate $\sigma^2$.
    The solid black lines represent the numerical occupation probabilities derived from Eq.~\eqref{eq:occProb}.}
    \label{fig:ReturnProbExpt}
\end{figure}

The particular form of matrices $\hat{\mathcal{U}}$ allows us to find a simple expression for the occupation probability $p_{j}(M)$ after $M$ steps of the state $|j\rangle$ at site $j$ in a slowly varying continuous limit (see Appendix~\ref{Appendix:master_edge}): 
\begin{equation}
p_j(M)\equiv\overline{|\langle j | \psi_M \rangle |^2}=\langle j |\overline{\rho_M} | j \rangle \label{eq:occProb}
\end{equation}
The site labeling in this case follows the natural order of sites with no sublattice distinction, that is, sites $j= 1, 2, 3, 4, 5, ...$ correspond to sublattice sites $\alpha_1, \beta_1, \alpha_2, \beta_2, \alpha_3, ...$, see Fig.~\ref{fig:ReturnProbExpt}(a).

For parameters $\theta_1=0.5\pi$, $\theta_2=0.0\pi$ and $\varphi=0.2\pi$, the lattice displays two bulk flat bands and an edge state on the leftmost site spectrally located at the $E=\pi$ gap. 
This edge state $|e_L\rangle$ is fully localized at site $L$ ($j=1$) at the left edge.
Its occupation probability, referred to as the edge-state return probability, can be expressed as
\begin{equation}
  p_{L}\left(M\right) =  \langle e_L |\overline{\rho_M} | e_L \rangle \approx p_{L}\left(0\right)e^{-M\Gamma_{+}}+\Gamma_{+}p_{L+1}\left(M\right).\label{eq:returnFinal}
\end{equation}

At short times, the initial condition $p_{L}\left(0\right)=1$ dominates, exponentially decreasing the occupation of the edge state.
At later times, the edge-state population is determined by the dynamics of the bulk modes, whose effect appears here through the term $p_{L+1}(M)$.
This behavior was anticipated in Ref.~\cite{rieder_localization_2018} where the decay of the edge state population was found to be exponential for dipersive bulk bands, and power-law for flat bands. 
Equation~\eqref{eq:returnFinal} allows understanding qualitatively this connection in an explicit form: the term $p_{L+1}\left(M\right)$ determines the subsequent decay and we expect its evolution to strongly depend on the bulk dispersion. 

The connection between the edge site $L$ dynamics and the bulk modes is experimentally studied in Fig.~\ref{fig:ReturnProbExpt}(b)-(d).
The considered lattice has $N = 44$ sites. 
The left edge site is initially populated ($p_{L}(0)=1$) and the return probability $ p_L(M) $ is measured and averaged over 100 noise realizations ($\tau_m$ sampled from a normal distribution with zero mean, $\overline{\tau}_{m}=0$ and $\overline{\tau_{m}\tau_{m^{\prime}}} = \sigma^{2}\delta_{m,m^{\prime}}$).
In the absence of noise (Fig.~\ref{fig:ReturnProbExpt}(b) and red dots in~(d)), the occupation probability of the edge site is close to one up to 70 steps.
At that time step, the amplifier gain in the fiber rings decreases and losses are no longer compensated.
In the presence of stroboscopic noise with amplitude $\sigma=0.12\pi$ the return probability (blue dots in Fig.~\ref{fig:ReturnProbExpt}(d)) decays exponentially at short times, followed by a crossover to a polynomial decay after roughly 8-10 steps.
This slowing-down arises from the dynamics of localized modes in the bulk flat bands.
This behavior is observed also for random noise (green dots). It closely matches the numerical return probabilities (blue solid lines obtained from the master equation, and it is significantly different from the exponential decay $e^{-\sigma^2 M}$ we would expect for the employed value of noise (gray dashed line).

Along with the decay of the occupation probability of the edge site, we observe an increase of the occupation of sites $j=2$ and $j=3$ (blue squares and triangles, respectively, in Fig.~\ref{fig:ReturnProbExpt}(d)) for the first 10-20 time steps and then a polynomial decrease.
Subsequent sites further in the bulk are expected to display similar dynamics, with a maximum occupation happening later in time step as we go deeper into the bulk.

The measured dynamics are well reproduced by the computation of the occupation probabilities given by the density matrix (Eq.~\eqref{eq:occProb} and Appendix~\ref{Appendix:master_edge}), shown as solid lines in Fig.~\ref{fig:ReturnProbExpt}(d).
The agreement is quantitative for sites $j=1,3$ and displays a factor of 2 discrepancy for site $j=2$, probably due to a calibration mismatch between the measured intensities of the $\alpha$ and $\beta$ rings in this specific experiment.

\section{Discussion}
In summary, we have studied the effect of noisy dynamics in DSQWs in a one-dimensional lattice using a density matrix formalism. 
We have shown that in DSQWs certain states do not show any decoherence under stroboscopic noise. 
Our results have direct relevance for architectures based on cascades of unitary operators with different interconnects, which can be prone to noise at timescales longer than the Floquet period.
This kind of noise might be important in architectures that use recycling schemes like the one in our experiments, in which a set of beam splitters is modified in time.
The study reported here reveals subspaces that can be profitable to build registers and to the experimental simulation of open quantum systems using quantum walks with enhanced robustness to temporal noise.

\begin{acknowledgments}
We thank I. C. Fulga for fruitful discussions at early stages of this work. This work was supported by the European Research Council grant EmergenTopo (865151), the French government through the Programme Investissement d'Avenir (I-SITE ULNE /ANR-16-IDEX-0004 ULNE) managed by the Agence Nationale de la Recherche, the Labex CEMPI (ANR-11-LABX-0007), and the region Hauts-de-France. It was partially funded by the CDP C2EMPI, as well as the French State under the France-2030 programme, the University of Lille, the Initiative of Excellence of the University of Lille, the European Metropolis of Lille for their funding and support of the R-CDP-24-004-C2EMPI project.
This project has received funding from the European Union’s Horizon 2020 research and innovation programme under the Marie Skłodowska-Curie grant agreement No 101108433.
A.G.L acknowledges support by MICIU/AEI/10.13039/501100011033 and by ERDF/EU to project PID2023-146531NA-I00. Also acknowledges support from CSIC Interdisciplinary Thematic Platform (PTI+) on Quantum Technologies (PTI-QTEP+).
\end{acknowledgments}

\textit{Data availability} —- The data that support the findings of this article are openly available~\cite{Data}.

\appendix

\section{Computation of the Floquet eigenvalues\label{sec:AppendixEigen}}

To compute the two-step Floquet eigenvalues in the noiseless case, we evolve one step Eq.~\ref{eq:BasicEq} to get:
\begin{align}
    \alpha_{n}^{m+2} =&	\left[-\alpha_{n}^{m}\sin\left(\theta_{m}\right)+i\beta_{n}^{m}\cos\left(\theta_{m}\right)\right]\sin\left(\theta_{m+1}\right)e^{i\varphi_{m+1}}\nonumber\\
	&+\left[\alpha_{n-2}^{m}\cos\left(\theta_{m}\right)+i\beta_{n-2}^{m}\sin\left(\theta_{m}\right)\right]\cos\left(\theta_{m+1}\right)e^{i(\varphi_{m}+\varphi_{m+1})}\label{eq:two-step1}\\
\beta_{n}^{m+2} =&	\left[i\alpha_{n}^{m}\cos\left(\theta_{m}\right)-\beta_{n}^{m}\sin\left(\theta_{m}\right)\right]\sin\left(\theta_{m+1}\right)e^{i\varphi_{m}}\nonumber\\
	&+\left[i\alpha_{n+2}^{m}\sin\left(\theta_{m}\right)+\beta_{n+2}^{m}\cos\left(\theta_{m}\right)\right]\cos\left(\theta_{m+1}\right).\label{eq:two-step2}
\end{align}

Given the periodicity of two steps in time and two sites in space, we search for eigenmodes solutions that follow the Floquet-Bloch ansatz:
\begin{flalign}
\left(\begin{array}{l}
\alpha_n^{m+1} \\
\beta_n^{m+1}\end{array}\right)&=\left(\begin{array}{l}\tilde{\alpha}(k) \\ 
\tilde{\beta}(k) \end{array}\right) e^{i \frac{E m}{2}} e^{i \frac{k n}{2}}. &&
\end{flalign}\\
By injecting this ansatz in Eqs.~\eqref{eq:two-step1} and~\eqref{eq:two-step2} and some manipulation, we get Eq.~\ref{eq:eigenvalues} for the Floquet eigenvalues.

\section{Master equation for the bulk and stroboscopic noise\label{sec:Appendix1}}
The time evolution of a state following a time-periodic protocol can be expressed in terms of the Floquet operator as $|\psi_{M}\rangle=\hat{U}_{F}^{M}|\psi_{0}\rangle$, where $M$ is the number of Floquet periods. 
However, in the presence of noise that is constant over a period, each Floquet operator at time $M$ becomes a function of a noise variable $\hat{U}_F(\tau_M)$.
In this case, in order to understand properties of the system that are robust to noise, it is useful to study noise averaged quantities, which require a density matrix treatment and the derivation of its corresponding master equation. Otherwise, non-Hermitian effective Hamiltonians will generally lead to non-conservation of the norm at long time due to the absence of jump operators. 
Defining the density matrix as $\hat{\rho}_M = |\psi_{M}\rangle\langle\psi_{M}|$, its expectation value over noise realizations $\overline{\hat{\rho}\left(M\right)}=\text{E}\left[\hat{\rho}\left(M\right)\right]$ is in general a complex calculation, with $\text{E}[\ldots]$ being the statistical average over a distribution. However, it highly simplifies if we assume uncorrelated noise, $\overline{\tau_{M}\tau_{L}}=\sigma^{2}\delta_{M,L}$, with zero average $\overline{\tau_{j}}=0$. This allows us to calculate the expectation value independently at each time-slice.
For the explicit calculation, it is useful to decompose each noisy Floquet operator in terms of different noise prefactors:
\begin{equation}
    \hat{U}_{F}\left(\tau_{M}\right)= \sum_{\mu}f_{\mu}\left(\tau_{M}\right)\hat{\mathcal{U}}_{\mu}
\end{equation}
This can be done in different ways, for example by expanding in Taylor series for small $\tau_j$ and then resuming the series, to obtain a non-perturbative expression.
Then, the noise average of the density matrix equation of motion can be expressed as:
\begin{align}
    \overline{\hat{\rho}_{M+1}} =& \text{E}\left[\hat{U}_{F}\left(\tau_{M+1}\right)\text{E}\left[\hat{\rho}_{M}\right]\hat{U}_{F}^{\dagger}\left(\tau_{M+1}\right)\right]\nonumber\\
    =& \sum_{\mu,\nu}F_{\mu,\nu}\left(\sigma^{2}\right)\hat{\mathcal{U}}_{\mu}\overline{\hat{\rho}_{M}}\hat{\mathcal{U}}_{\nu}^{\dagger}
\end{align}
with $F_{\mu,\nu}\left(\sigma^{2}\right)=\text{E}\left[f_{\mu}\left(\tau_{M}\right)f_{\nu}\left(\tau_{M}\right)\right]$.

For the case of bulk dynamics, we can express the evolution operator in momentum space, and write the evolution for each step $m$ of the protocol as a 2-dimensional matrix:
\begin{equation}
    \hat{U}_{m}\left(k\right)=\left(\begin{array}{cc}
\cos\left(\theta_{m}\right)e^{i\left(\varphi_{m}-\frac{k}{2}\right)} & i\sin\left(\theta_{m}\right)e^{i\left(\varphi_{m}-\frac{k}{2}\right)}\\
ie^{\frac{ik}{2}}\sin\left(\theta_{m}\right) & e^{\frac{ik}{2}}\cos\left(\theta_{m}\right)
\end{array}\right)
\end{equation}
The noisy Floquet operator for a 2-steps protocol at time $M$, $\hat{U}_{F}\left(k,\tau_{M}\right)=\hat{U}_{2}\left(k,\tau_M\right)\hat{U}_{1}\left(k,\tau_M\right)$, defined by the set of parameters $\theta_{1}, \theta_2$ and $\varphi\}$, is then obtained by choosing $\theta_{1,2}\to\theta_{1,2}+\tau_{M}$, where again $M$ indicates the period number. Using trigonometric relations, it can be decomposed in terms of three matrices only:
\begin{equation}
    \hat{U}_{F}\left(k,\tau_{M}\right)=\hat{U}_{F}\left(k\right)+f_{+}\left(\tau_{M}\right)\hat{\mathcal{U}}_{+}(k)+f_{-}\left(\tau_{M}\right)\hat{\mathcal{U}}_{-}(k)
\end{equation} 
with $f_{-}\left(\tau_{M}\right)= -\sin\left(\tau_{M}\right)\cos\left(\tau_{M}\right)$, $f_{+}\left(\tau_{M}\right)=-\sin^{2}\left(\tau_{M}\right)$ and the 2-dimensional matrices being:
\begin{align}
    \hat{\mathcal{U}}_{-}(k) &= \left(\begin{array}{cc}
s_{-}(k,\varphi)\sin\left(\theta_{+}\right) & -is_{-}(k,\varphi)\cos\left(\theta_{+}\right)\\
-is_{+}(k,\varphi)\cos\left(\theta_{+}\right) & s_{+}(k,\varphi)\sin\left(\theta_{+}\right)
\end{array}\right)\nonumber\\
\hat{\mathcal{U}}_{+} (k) &= \left(\begin{array}{cc}
s_{-}(k,\varphi)\cos\left(\theta_{+}\right) & is_{-}(k,\varphi)\sin\left(\theta_{+}\right)\\
is_{+}(k,\varphi)\sin\left(\theta_{+}\right) & s_{+}(k,\varphi)\cos\left(\theta_{+}\right)
\end{array}\right)\nonumber
\end{align}
with $\theta_+=\theta_1+\theta_2$ and $s_\pm (k,\varphi)= e^{\pm ik}+e^{\pm i\varphi}$.
Finally, performing the noise average for Gaussian noise (other choices are possible), we arrive at the master equation for the bulk dynamics at a particular $k$:
\begin{widetext}
\begin{align}
    \overline{\hat{\rho}_{M+1}} =& \hat{U}_{F}\overline{\hat{\rho}_M}\hat{U}_{F}^{\dagger}-\frac{1}{2}\left(1-e^{-2\sigma^{2}}\right)\left(\hat{\mathcal{U}}_{+}\overline{\hat{\rho}_M}\hat{U}_{F}^{\dagger}+\hat{U}_{F}\overline{\hat{\rho}_M}\hat{\mathcal{U}}_{+}^{\dagger}\right)\\
    &+\frac{1}{8}\left(3+e^{-8\sigma^{2}}-4e^{-2\sigma^{2}}\right)\hat{\mathcal{U}}_{+}\overline{\hat{\rho}_M}\hat{\mathcal{U}}_{+}^{\dagger}+\frac{1}{8}\left(1-e^{-8\sigma^{2}}\right)\hat{\mathcal{U}}_{-}\overline{\hat{\rho}_M}\hat{\mathcal{U}}_{-}^{\dagger}
\end{align}
where we have used that for Gaussian noise:
\begin{align}
    \text{E}\left[f_{+}\left(\tau\right)\right]&=-\frac{1}{2}\left(1-e^{-2\sigma^{2}}\right),\ \text{E}\left[f_{-}\left(\tau\right)\right]=0,\ \text{E}\left[f_{-}\left(\tau\right)f_{+}\left(\tau\right)\right]=0\\
    \text{E}\left[f_{+}^{2}\left(\tau\right)\right]&=\frac{1}{8}\left(3+e^{-8\sigma^{2}}-4e^{-2\sigma^{2}}\right),\ \text{E}\left[f_{-}^{2}\left(\tau\right)\right]=\frac{1}{8}\left(1-e^{-8\sigma^{2}}\right)
\end{align}
\end{widetext}
Notice that the first term in the master equation corresponds to the noiseless $(\sigma\rightarrow 0$) evolution under the Floquet operator, while the others describe the decoherence introduced by the noise. Importantly, as the entries in the matrices $\hat{\mathcal{U}}_\pm$ are proportional to $e^{\pm ik}+e^{\pm i\varphi}$, one can see that for $k=\varphi+(2p+1)\pi$, for all $p\in\mathbb{Z}$, they vanish. This is the origin of the tunable decoherence free subspaces.
Also, it is important to notice that, for small $\sigma$, the term proportional to $3+e^{-8\sigma^{2}}-4e^{-2\sigma^{2}}\sim\mathcal{O}(\sigma^4)$, indicating that it is a correction that goes beyond the Lindblad form that is obtained at second order in $\sigma$~\cite{sieberer_statistical_2018,rieder_localization_2018}.

\section{Master equation for the bulk and random noise}
For non-stroboscopic noise we must consider the single step evolution $\hat{U}_{m}\left(\tau_{m}\right)$, with random variable $\tau_{m}$ at each step, instead of the Floquet operator (which is no longer well-defined). The single step evolution operator for the bulk dynamics reads:
\begin{equation}
    \hat{U}_{m}\left(k\right)= \left(\begin{array}{cc}
\cos\left(\theta_{m}\right)e^{i\left(\varphi_{m}-\frac{k}{2}\right)} & i\sin\left(\theta_{m}\right)e^{i\left(\varphi_{m}-\frac{k}{2}\right)}\\
ie^{\frac{ik}{2}}\sin\left(\theta_{m}\right) & e^{\frac{ik}{2}}\cos\left(\theta_{m}\right)
\end{array}\right)
\end{equation}
Introducing the noise via $\theta_m\to\theta_m+\tau_m$ and factorizing, we can write the noise evolution operator as:
\begin{equation}
    \hat{U}_{m}\left(k,\tau_{m}\right) = \cos\left(\tau_{m}\right)\hat{U}_{m}(k)+\sin\left(\tau_{m}\right)\hat{U}_{m}^\prime(k),
\end{equation}
 with $\hat{U}_{m}(k)=\left.\hat{U}_{m}(k,\tau_m)\right|_{\tau_m=0}$ and $\hat{U}_{m}^\prime(k)=\left.\partial_{\tau_m}\hat{U}_{m}(k,\sigma)\right|_{\tau_m=0}$.
 
 Finally, calculating the expectation value over noise realizations of the density matrix we arrive at the following master equation for a particular $k$ after a single step:
\begin{equation}
    \overline{\hat{\rho}_{m+1}} = \frac{1+e^{-2\sigma^{2}}}{2}\hat{U}_{1}\overline{\hat{\rho}_{m}}\hat{U}_{1}^{\dagger} + \frac{1-e^{-2\sigma^{2}}}{2} \hat{U}_{1}^{\prime}\overline{\hat{\rho}_{m}}\hat{U}_{1}^{\prime\dagger}.
\end{equation}
 For the case of interest, with a 2-step protocol, this expression can be iterated to an additional time-step, to write the dynamics after an approximate single Floquet protocol, by:
 \begin{widetext}
 \begin{equation}
     \overline{\hat{\rho}_{m+2}}=\left(\frac{1+e^{-2\sigma^{2}}}{2}\right)^{2}\hat{U}_{F}\overline{\hat{\rho}_{m}}\hat{U}_{F}^{\dagger}+\frac{1-e^{-4\sigma^{2}}}{4}\left(\hat{U}_{2}\hat{U}_{1}^{\prime}\overline{\hat{\rho}_{m}}\hat{U}_{1}^{\prime\dagger}\hat{U}_{2}^{\dagger}+\hat{U}_{2}^{\prime}\hat{U}_{1}\overline{\hat{\rho}_{m}}\hat{U}_{1}^{\dagger}\hat{U}_{2}^{\prime\dagger}\right)+\left(\frac{1-e^{-2\sigma^{2}}}{2}\right)^{2}\hat{U}_{2}^{\prime}\hat{U}_{1}^{\prime}\overline{\hat{\rho}_{m}}\hat{U}_{1}^{\prime\dagger}\hat{U}_{2}^{\prime\dagger}
 \end{equation}
 \end{widetext}
 where it is clear that the first contribution produces the noiseless dynamics, and that it is always affected if $\sigma\neq 0$, unlike for the case of stroboscopic noise.

\section{Master equation for the edge states and stroboscopic noise}\label{Appendix:master_edge}
For the edge state dynamics under stroboscopic noise, the calculation of the master equation is identical to that of the bulk under stroboscopic noise, but the matrices are of dimension $2N\times2N$ due to the real space formulation. The Floquet operator can be written in terms of $2\times2$ blocks as:
\begin{equation}
    \hat{U}_F = \left(\begin{array}{cccccc}
\hat{U}_{L} & \hat{U}_{+} & 0 & 0 & 0 & \udots\\
\hat{U}_{-} & \hat{U}_{0} & \hat{U}_{+} & 0 & \ddots & 0\\
0 & \hat{U}_{-} & \hat{U}_{0} & \ddots & 0 & 0\\
0 & 0 & \ddots & \hat{U}_{0} & \hat{U}_{+} & 0\\
0 & \ddots & 0 & \hat{U}_{-} & \hat{U}_{0} & \hat{U}_{+}\\
\udots & 0 & 0 & 0 & \hat{U}_{-} & \hat{U}_{R}
\end{array}\right)\label{eq:FloquetOperatorOBC}
\end{equation}

where each block is defined as
\begin{align}
    \hat{U}_{0}=& \sin\left(\theta_{2}\right)\left(\begin{array}{cc}
-e^{-i\varphi}\sin\left(\theta_{1}\right) & ie^{-i\varphi}\cos\left(\theta_{1}\right)\\
ie^{i\varphi}\cos\left(\theta_{1}\right) & -e^{i\varphi}\sin\left(\theta_{1}\right)
\end{array}\right)\nonumber\\
\hat{U}_{+}=& \cos\left(\theta_{2}\right)\left(\begin{array}{cc}
0 & 0\\
i\sin\left(\theta_{1}\right) & \cos\left(\theta_{1}\right)
\end{array}\right)\nonumber\\
\hat{U}_{-}=& \cos\left(\theta_{2}\right)\left(\begin{array}{cc}
\cos\left(\theta_{1}\right) & i\sin\left(\theta_{1}\right)\\
0 & 0
\end{array}\right)\nonumber\\
\hat{U}_{L}=& \left(\begin{array}{cc}
-e^{-i\varphi}\sin\left(\theta_{1}\right) & ie^{-i\varphi}\cos\left(\theta_{1}\right)\\
ie^{i\varphi}\cos\left(\theta_{1}\right)\sin\left(\theta_{2}\right) & -e^{i\varphi}\sin\left(\theta_{1}\right)\sin\left(\theta_{2}\right)
\end{array}\right)\nonumber\\
\hat{U}_{R}=& \left(\begin{array}{cc}
-e^{-i\varphi}\sin\left(\theta_{1}\right)\sin\left(\theta_{2}\right) & ie^{-i\varphi}\cos\left(\theta_{1}\right)\sin\left(\theta_{2}\right)\\
ie^{i\varphi}\cos\left(\theta_{1}\right) & -e^{i\varphi}\sin\left(\theta_{1}\right)
\end{array}\right)\nonumber
\end{align}
In particular, the decomposition of the noisy Floquet operator in real space results in five different contributions, which can be obtained straightforwardly from Eq.~\eqref{eq:FloquetOperatorOBC} by the use of the trigonometric identities $\cos(\theta+\tau)=\cos(\theta)\cos(\tau)-\sin(\theta)\sin(\tau)$ and $\sin(\theta+\tau)=\cos(\theta)\sin(\tau)+\sin(\theta)\cos(\tau)$:
\begin{align}
    \hat{U}_{F}\left(\tau_{M}\right)=& \hat{\mathcal{U}}_{0}+ \cos\left(\tau_{M}\right)\hat{\mathcal{U}}_{c}+ \sin\left(\tau_{M}\right)\hat{\mathcal{U}}_{s}+ \cos^{2}\left(\tau_{M}\right)\hat{\mathcal{U}}_{cc}\nonumber\\
    &+\sin^{2}\left(\tau_{M}\right)\hat{\mathcal{U}}_{ss}+ \sin\left(\tau_{M}\right)\cos\left(\tau_{M}\right)\hat{\mathcal{U}}_{sc}
\end{align}
Here, the matrices $\hat{\mathcal{U}}_{c}$, $\hat{\mathcal{U}}_{s}$, $\hat{\mathcal{U}}_{cc}$ etc. are obtained by identifying common prefactors $\cos(\tau_M)$, $\sin(\tau_M)$, $\cos^2(\tau_M)$ etc. and grouping them together.
Interestingly, the matrices $\hat{\mathcal{U}}_{c}$ and $\hat{\mathcal{U}}_{s}$ only couple the edge with its nearest neighbor site.
This indicates that these noise processes act only on the edge states. In contrast, the matrices $\hat{\mathcal{U}}_{cc}$, $\hat{\mathcal{U}}_{ss}$ and $\hat{\mathcal{U}}_{sc}$ only couple neighboring bulk sites, leaving the edges unaffected. This separation will allows us to physically interpret the different terms in the master equation below.

Finally, the master equation is obtained after a noise average, which for the case of Gaussian noise, reads:
\begin{widetext}
\begin{align}
    \overline{\hat{\rho}_{M+1}}=& \hat{\mathcal{U}}_{0}\overline{\hat{\rho}_{M}}\hat{\mathcal{U}}_{0}^{\dagger}+e^{-\frac{\sigma^{2}}{2}}\left(\hat{\mathcal{U}}_{0}\overline{\hat{\rho}_{M}}\hat{\mathcal{U}}_{c}^{\dagger}+\hat{\mathcal{U}}_{c}\overline{\hat{\rho}_{M}}\hat{\mathcal{U}}_{0}^{\dagger}\right)+\frac{1}{2}\left(1+e^{-2\sigma^{2}}\right)\left(\hat{\mathcal{U}}_{c}\overline{\hat{\rho}_{M}}\hat{\mathcal{U}}_{c}^{\dagger}+\hat{\mathcal{U}}_{0}\overline{\hat{\rho}_{M}}\hat{\mathcal{U}}_{cc}^{\dagger}+\hat{\mathcal{U}}_{cc}\overline{\hat{\rho}_{M}}\hat{\mathcal{U}}_{0}^{\dagger}\right)\nonumber\\
    &+\frac{1}{2}\left(1-e^{-2\sigma^{2}}\right)\left(\hat{\mathcal{U}}_{s}\overline{\hat{\rho}_{M}}\hat{\mathcal{U}}_{s}^{\dagger}+\hat{\mathcal{U}}_{0}\overline{\hat{\rho}_{M}}\hat{\mathcal{U}}_{ss}^{\dagger}+\hat{\mathcal{U}}_{ss}\overline{\hat{\rho}_{M}}\hat{\mathcal{U}}_{0}^{\dagger}\right) +\frac{1}{4}\left(3e^{-\frac{1}{2}\sigma^{2}}+e^{-\frac{9}{2}\sigma^{2}}\right)\left(\hat{\mathcal{U}}_{c}\overline{\hat{\rho}_{M}}\hat{\mathcal{U}}_{cc}^{\dagger}+\hat{\mathcal{U}}_{cc}\overline{\hat{\rho}_{M}}\hat{\mathcal{U}}_{c}^{\dagger}\right) \nonumber\\
    &+\frac{1}{4}\left(e^{-\frac{1}{2}\sigma^{2}}-e^{-\frac{9}{2}\sigma^{2}}\right)\left(\hat{\mathcal{U}}_{c}\overline{\hat{\rho}_{M}}\hat{\mathcal{U}}_{ss}^{\dagger}+\hat{\mathcal{U}}_{ss}\overline{\hat{\rho}_{M}}\hat{\mathcal{U}}_{c}^{\dagger}+\hat{\mathcal{U}}_{s}\overline{\hat{\rho}_{M}}\hat{\mathcal{U}}_{sc}^{\dagger}+\hat{\mathcal{U}}_{sc}\overline{\hat{\rho}_{M}}\hat{\mathcal{U}}_{s}^{\dagger}\right) \nonumber\\
    &+\frac{1}{8}\left(1-e^{-8\sigma^{2}}\right)\left(\hat{\mathcal{U}}_{sc}\overline{\hat{\rho}_{M}}\hat{\mathcal{U}}_{sc}^{\dagger}+\hat{\mathcal{U}}_{ss}\overline{\hat{\rho}_{M}}\hat{\mathcal{U}}_{cc}^{\dagger}+\hat{\mathcal{U}}_{cc}\overline{\hat{\rho}_{M}}\hat{\mathcal{U}}_{ss}^{\dagger}\right)\nonumber\\
    &+\frac{1}{8}\left(3+4e^{-2\sigma^{2}}+e^{-8\sigma^{2}}\right)\hat{\mathcal{U}}_{cc}\overline{\hat{\rho}_{M}}\hat{\mathcal{U}}_{cc}^{\dagger}+\frac{1}{8}\left(3-4e^{-2\sigma^{2}}+e^{-8\sigma^{2}}\right)\hat{\mathcal{U}}_{ss}\overline{\hat{\rho}_{M}}\hat{\mathcal{U}}_{ss}^{\dagger}\label{eq:MasterEqEdge}
\end{align}
\end{widetext}

\section{Edge state dynamics under stroboscopic noise}
To study the noisy return probability of an edge state after $M$ periods, $p_\text{L}(M)=\langle e_{L}|\overline{\hat{\rho}_{M}}|e_{L}\rangle$, it is useful to consider the fully dimerized case $\theta_1=\pi/2$ and $\theta_2=0$. In this situation the system corresponds to the flat band case and has a fully localized edge state $|e_\text{L}\rangle=(1,0,0,\ldots)$, which we consider the initial state for the density matrix $\hat{\rho}_0=|e_\text{L}\rangle \langle e_\text{L}|$.
Importantly, one can check that several noise operators from the master equation, Eq.~\eqref{eq:MasterEqEdge}, do not affect the edge state:
\begin{equation}
    \hat{\mathcal{U}}_{0}^{\dagger}|e_{L}\rangle=0,\ \hat{\mathcal{U}}_{cc}^{\dagger}|e_{L}\rangle=0,\ \hat{\mathcal{U}}_{ss}^{\dagger}|e_{L}\rangle=0\text{ and }\hat{\mathcal{U}}_{sc}^{\dagger}|e_{L}\rangle=0.
\end{equation}
This massively simplifies the calculation of the master equation for the population of the edge site to:
\begin{equation}
    p_\text{L}(M+1)= \frac{1+e^{-2\sigma^{2}}}{2} p_\text{L}(M)+\frac{1-e^{-2\sigma^{2}}}{2} p_{\text{L}+1}(M)
\end{equation}
This is a recurrence equation for the population, which can be turned into the following differential equation for the continuum time variable $t$:
\begin{equation}
    \partial_{t}p_{L}\left(t\right)\approx -\Gamma_{+}\left[p_{L}\left(t\right)-p_{L+1}\left(t\right)\right]
\end{equation}
with $\Gamma_{+}\equal \frac{1-e^{-2\sigma^{2}}}{2}$. This simple equation allows to correctly estimate the short-time and long-time behavior of the return probability. Its formal solution is:
\begin{equation}
    p_{L}\left(t\right)=p_{L}\left(0\right)e^{-\Gamma_{+} t}+ \Gamma_{+}\int_{0}^{t}e^{-\Gamma_{+}\left(t-s\right)}p_{L+1}\left(s\right)ds
\end{equation}
Notice that for an initial condition with the edge state occupied, $p_{\text{L}}(0)=1$, the dynamics is controlled by the first term, which predicts an exponential decay with $\Gamma_{+}\approx\sigma^2$ for small $\sigma$, as predicted from the Lindblad master equation~\cite{rieder_localization_2018}.
However, at long time the first term goes to zero and the second term dominates, which is a kernel that convolutes the exponential decay with the population at the neighboring site.
If we apply the change of variable $u= t-s$, the previous equation becomes:
\begin{equation}
    p_{L}\left(t\right)= p_{L}\left(0\right)e^{-\Gamma_{+} t}+\Gamma_{+}\int_{0}^{t}e^{-\Gamma_{+} u}p_{L+1}\left(t-u\right)du
\end{equation}
where we can see that due to the fast decay introduced by the exponential, the kernel mostly contributes when $u=0$ or equivalently $t=s$. Hence, we can approximate the solution by $p_{L}\left(t\right)\approx p_{L}\left(0\right)e^{-\Gamma_{+} t}+\Gamma_{+} p_{L+1}\left(t\right)$, when the dynamics of the population at bulk sites is slow compared with the decay, which is the case for flat bands or disordered systems.
Therefore, we can see that our solution predicts an exponential decay at short times that is taken over by the dynamics of the population at the bulk sites.

\section{Experimental set-up}
The experimental setup employs a continuous wave (CW) single-frequency laser source (Koheras MIKRO, NKT Photonics) operating at 1550 nm with a maximum output power of 40 mW and a linewidth $<$ 0.1 kHz. 
The laser output is split equally using a 50/50 beam splitter, with one part serving as input to a local oscillator and the other being modulated into 1.4 ns pulses via an electro-optical modulator (EOM, iXblue MXER-LN-10), which is controlled by an arbitrary waveform generator (AWG 7000B, Tektronix). 
To reduce residual laser light entering the ring, an acoustic optical modulator (AOM, AA Opto-electronic MT110-IIR30-Fio-PM0.5) with an extinction power of -70 dB is incorporated. 
The AOM is shaped in a gate centered in time at the pulse generated by the preceding EOM. 
The prepared injection signal is then introduced into the long $\alpha$ ring through a 70/30 beamsplitter.

The pulse evolution follows a split step walk. 
The two fiber rings $\alpha$ and $\beta$ are coupled via a high-bandwidth 40 GHz electronically controlled variable beamsplitter (EOSpace AX-2x2-0MSS-20). 
Each ring has an Erbium-doped fiber amplifier (EDFA, Keopsys CEFA-C-HG) and an optical variable attenuator (VOA, Agiltron) which are used to finely compensate for round trip losses.

A 90/10 beamsplitter within each ring extracts light for measurement. 
To access both amplitude and phase information of sublattices $\alpha_n^m$ and $\beta_n^m$, a heterodyne measurement technique is employed. 
This involves beating the wavefield extracted from the double rings with a local oscillator reference field.
This field is derived from the laser used to inject the initial pulse and is frequency-shifted by 3 GHz using an electro-optic modulator.
The beating interference between the signal and the local oscillator is converted to electrical signals using a fast photo-diode (Thorlabs DET08CFC) operating at 5 GHz. 
These signals are then captured and analyzed using a high-performance oscilloscope (Tektronix MSO64) featuring a 6 GHz bandwidth, 10-bit vertical resolution, 25 GS/s sampling rate, and a memory record length of 62.5 Mpts corresponding to 2.5 ms, enabling very detailed signal analysis of the beating.

\section{Data processing}\label{Appendix:data_proc}
In this section we explain how to extract complex valued spatio-temporal dynamics and band diagram from measured real valued signal intensity.

\begin{figure}
    \centering
    \includegraphics[width=\columnwidth]{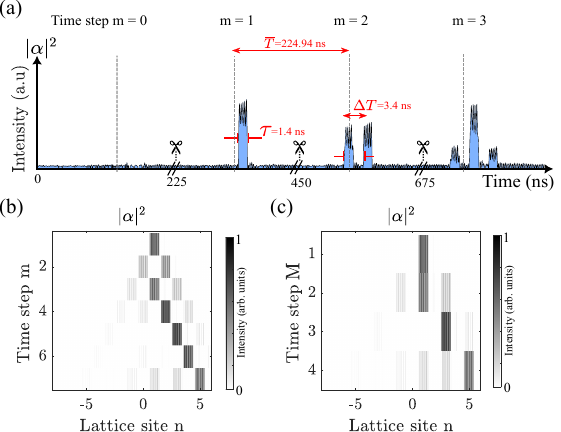}
    \caption{
    (a) Zoom on the first time steps of the measured time trace of the signal intensity at the output of the $\alpha$ fiber loop. 
    (b) Spatio-temporal diagram of the $\alpha$ ring reconstructed from the time trace shown in (a).
    (c) Corresponding stroboscopic spatio-temporal diagram obtained by sampling only odd time steps from (b).
    }
    \label{fig:SpatioTempDiagSupp}
\end{figure}

Data is collected by recording the output intensity from both fiber loops using an oscilloscope. 
The resulting signal displays groups of pulses separated by the average round-trip time of $\Bar{T}$ = 224.94 ns, with pulses within each group spaced $\Delta T$ = 3.4 ns apart due to the length difference between the two fiber loops as shown in  Fig.~\ref{fig:SpatioTempDiagSupp}.(a). 
This time trace is then segmented and arranged into a spatio-temporal diagram (Fig.~\ref{fig:SpatioTempDiagSupp}.(b)) for further analysis. 
Crucial phase information is retrieved by subjecting the signal to optical heterodyne measurement with a reference continuous wave laser, frequency-shifted by about 3 GHz, producing observable fringes in the recorded signal.

\begin{figure}
    \centering
    \includegraphics[width=\columnwidth]{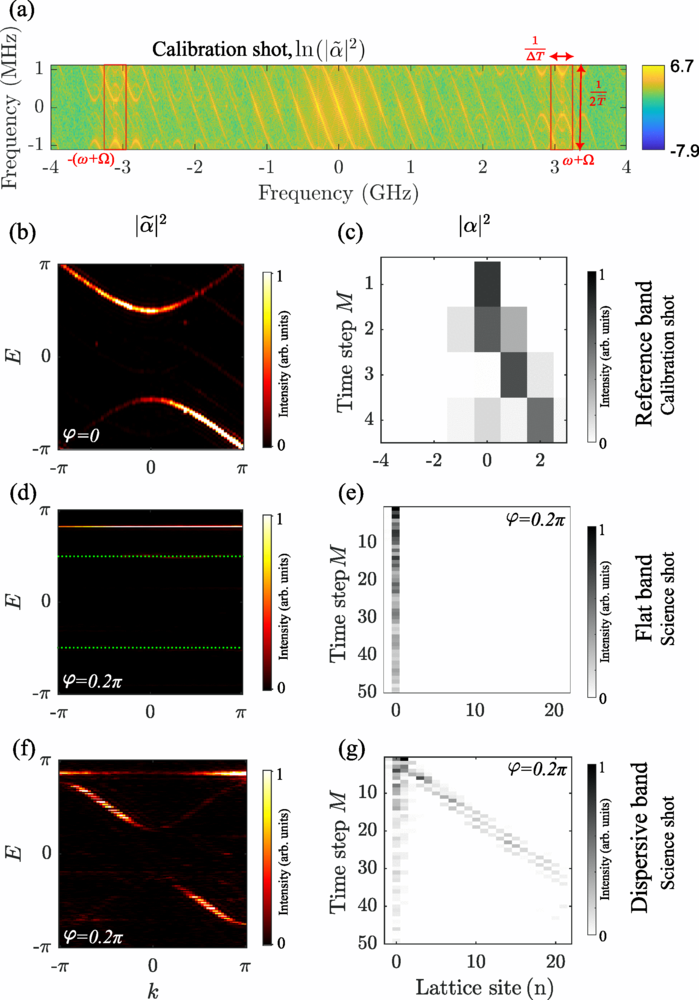}
    \caption{
    (a) Two-dimensional Fourier transform (2D-FT) of the stroboscopic spatio-temporal diagram of the $\alpha$ (Fig.~\ref{fig:SpatioTempDiagSupp}(c)).
    (b) Measured bands in one Brillouin zone for the $\alpha$ after zooming the region at $\omega + \Omega$ frequencies in panel (a) for the calibration shot ($\theta_1=0.25\pi$, $\theta_2=0.25\pi$, $\varphi=0.0\pi$).
    (c) Complex-valued spatio-temporal dynamics obtained by applying a two-dimensional inverse Fourier transform (2D-IFT) to the calibration band data in (b).
    (d) Measured band structure for the science shot of flat band ($\theta_1=0.5\pi$, $\theta_2=0.0\pi$, $\varphi=0.2\pi$) within the same Brillouin zone. Green dashed lines show the bulk bands.
    (e) Complex-valued spatio-temporal dynamics derived from the science pulse band data in (d) via 2D-IFT for flat band.
    (f) Measured band structure for the science shot of dispersive band ($\theta_1=0.25\pi$, $\theta_2=0.0\pi$, $\varphi=0.2\pi$) within the same Brillouin zone.
    (g) Complex-valued spatio-temporal dynamics derived from the science pulse band data in (f) via 2D-IFT for dispersive band.
    Panels (c), (e) and (g) display the initial time steps of the spatio-temporal evolution, with the full diagrams spanning approximately 40 stroboscopic steps for the calibration shot and 220 steps for the science shot.
    }
    \label{fig:BandsSupp}
\end{figure}

The interference between the local oscillator and the signal evolving in the rings contains phase information relevant to the measurement of the band structure (see  the beating signal on top of each pulse in Fig.~\ref{fig:SpatioTempDiagSupp}.(b)). 
The band structure is reconstructed by performing a numerical two-dimensional Fourier transform (2D-FT) on the stroboscopic spatio-temporal diagram of each ring at time steps corresponding to integer Floquet periods ($m = 1,3,5,\cdots$) as shown in Fig.~\ref{fig:SpatioTempDiagSupp}.(c). 
This yields periodic eigenvalue bands spanning about 25 GHz in the quasimomentum direction this value is fixed by the time resolution of the oscilloscope that records the time trace) and 2.24 MHz in the quasienergy direction, see Fig.~\ref{fig:BandsSupp}(a).
We focus on a single Brillouin zone at around a frequency of 3GHz as shown in Fig.~\ref{fig:BandsSupp}(a).
The vertical and horizontal axis of the dispersion are then relabeled to span the full spectral Brillouin zone both in quasienergy $E$ and quasimomentum $k$, spanning both from $-\pi$ to $\pi$ (Fig.~\ref{fig:BandsSupp}(b)).

Environmental factors can cause fluctuations in fiber length, resulting in shifts of the band structure. 
To diminish these fluctuations we use a protocol using piezos to lock the lengths of the rings. 
Even after this compensation we still have slight shifts in band structure. 
To account for these minor shifts, the experimental setup employs a double measurement protocol. A first pulse is injected and evolves in a nominally noiseless lattice following a simple model ($\theta_m =\pi/4$). The measured evolution is used to calibrate the evolution of a second pulse that now follows the lattice dynamics of interest (arbitrary values of $\theta_m$ and different noise realizations)

This method utilizes two consecutive $\tau$ = 1.4 ns pulses: a calibration pulse and a science pulse. 
The calibration pulse, enters the ring and evolves in time with a constant splitting of like 50/50 in the variable beamsplitter and no phase modulation.
The spatio-temporal evolution dynamics produces a well-known reference band structure as shown in Fig.~\ref{fig:BandsSupp}(b), where $\theta_1=0.25\pi, \theta_2=0.25\pi$ and $\varphi=0$. 
In contrast, the science pulse implements the experimental system of interest, featuring controlled coupling values of the variable beamsplitter and phase modulation of phase modulator.
The calibration pulse's band structure is compared to its theoretical model, allowing for the measurement of horizontal and vertical shifts. 
These measurements are then used to calibrate the axes, which remain valid for the subsequent science pulse measurement. And it allows us to calculate the science shot band structure as shown in Fig.~\ref{fig:BandsSupp}(d), where $\theta_1=0.5\pi, \theta_2=0$ and $\varphi=0.2\pi$. 

To recover the complex-valued spatio-temporal evolution, an inverse two-dimensional Fourier transform (2D-IFT) is applied to the band diagrams from both the calibration and science pulse measurements, as shown in Fig.~\ref{fig:BandsSupp}(c) and (e). This transformation converts observed intensity profiles (see Fig.~\ref{fig:SpatioTempDiagSupp}(c)) into detailed complex-valued spatio-temporal data. Notably, since the initial square pulse configuration introduces a random global phase with each initialization, this phase is subtracted from all measured dynamics to maintain consistency across shots.

To extract the experimental return probability to the topological edge state, we project the measured spatiotemporal dynamics of the full wavefunction $| \psi_M \rangle$ constructed from both $\alpha$ and $\beta$ sublattices into the edge state computed from the diagonalization of the Floquet operator of a finite size lattice Eq.~\eqref{eq:FloquetOperatorOBC}. 
We then proceed to the averaging over noise realizations.
Figure~\ref{fig:BandsSupp}(e) and (g) present examples of the measured dynamics in the $\alpha$ sublattice during the initial time steps following an edge site excitation, for the flat band and dispersive band cases, respectively.

\begin{figure}
    \centering
    \includegraphics[width=\columnwidth]{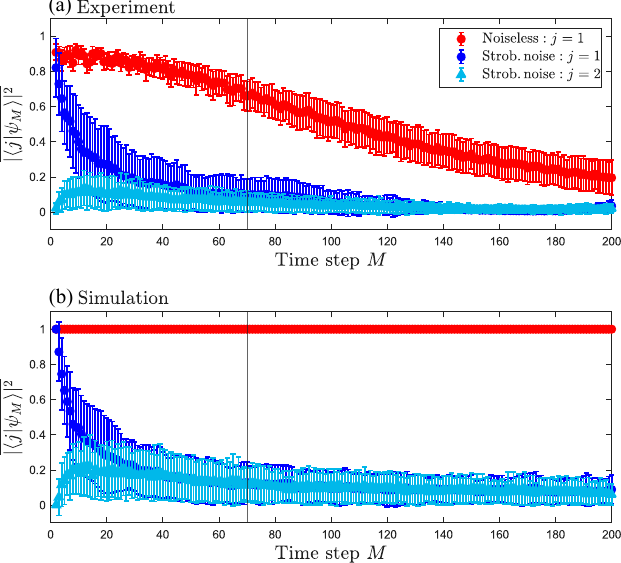}
    \caption{(a) Experimentally measured and (b) simulated averaged occupation probabilities, $\overline{|\langle j | \psi_M \rangle|^2} $ as a function of time step $M$ following initial excitation of the left edge site $j=1$, for finite lattice with flat band parameters $\theta_1=0.5\pi$, $\theta_2=0.0\pi$, $\varphi=0.2\pi$ and $N=44$ number of sites. 
    Edge state dynamics are shown for the noiseless case ($\sigma=0\pi$) in red dots and for stroboscopic noise ($\sigma=0.12\pi$) in blue dots, averaged over 100 realizations.
    Blue triangles show the evolution of bulk sites occupation.
    Error bars denote the standard deviation across realizations.}
    \label{fig:EdgeStateError}
\end{figure}

\begin{figure}
    \centering
    \includegraphics[width=0.8\columnwidth]{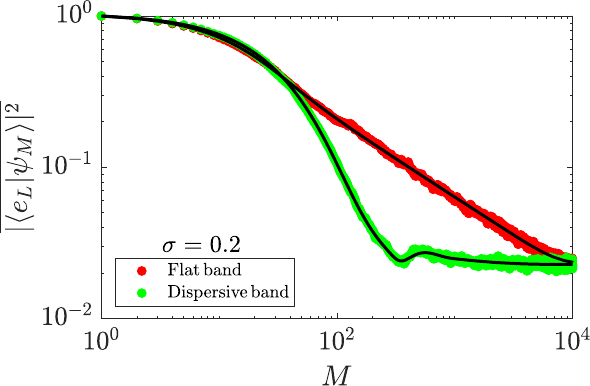}
    \caption{Numerically computed edge state return probability $\overline{|\langle e_L | \psi_M \rangle|^2} $ as a function of step $M$ in a lattice with $N=44$ sites and averaged over 1000 independent realizations. 
    In the flat band ($\theta_1=0.5\pi$, $\theta_2=0.0\pi$, $\varphi=0.2\pi$) with stroboscopic noise $\sigma=0.2$, the red dots (numerically calculated) show the edge state return probability exhibiting an initial exponential decay followed by a polynomial decay at intermediate times.
    In contrast, the dispersive band ($\theta_1=0.45\pi$, $\theta_2=0.0\pi$, $\varphi=0.2\pi$) with same noise the dynamics given by green dots (numerically calculated) mainly shows exponential decay with finite-size saturation at long times.
    The solid black lines represent the numerical edge state return probabilities derived from Eq.~(12) in the main text.}
    \label{fig:ReturnProbSimSupp}
\end{figure}

In the main text, Fig.~4 shows the edge state return probability for the flat band case in log--log scale.
Figure~\ref{fig:EdgeStateError}(a) shows the same data over the full 200 recorded time steps in linear scale, with error bars representing the standard deviation across 100 realizations.
The corresponding simulated results are shown in Fig.~\ref{fig:EdgeStateError}(b) for the same parameters.
Ideally edge state return probability for the noiseless case should be equal to 1 at all time steps (as in simulated results).
The observed decay and increasing standard deviation arise from experimental errors (mostly from fibre length variations, electronic noise and data processing), which cause variations in the extracted amplitude and phase for each realization. 
When averaged over 100 realizations, these errors accumulate, producing the decay and spread visible in Fig.~\ref{fig:EdgeStateError}(a).
When engineered stroboscopic noise is introduced in the couplers in the experiment  ($\sigma = 0.12\pi$), additional broadening is observed due to variations between noise realizations. 
Over longer time steps, the signal-to-noise ratio gradually degrades, reducing measurement reliability. 
For this reason, the main text displays only the initial 70 steps (corresponding to about a 30\% signal reduction) in Fig.~\ref{fig:ReturnProbExpt}. 
Note that the relative fluctuation in the numerical case for site $j=1$ and $j=2$ reaches roughly $100\%$ after 20 steps. 
This is due to different exploration of various dynamics due to the noise variation from one realization to another. 
We observe similar relative fluctuation in the experiment.

\section{Edge state dynamics with dispersive bulk bands}\label{Appendix:edge_dispersive}
Let us know provide a complementary analysis of the edge state dynamics in the case of a model realising a bulk with dispersive bands.
A numerical study of this situation using a continuous time model was discussed in Ref.~\cite{rieder_localization_2018}. 
Measuring the edge state return probability for a dispersive band model is more challenging than doing it for a flat band. 
The reason is that in a flat band, the edge sates are fully localized on a single edge site, which can be simply excited and measured. 
In a dispersive band, the complex-valued edge state extends over multiple sites, causing an excitation at the edge site to simultaneously excite edge and bulk modes, as illustrated in Fig.~\ref{fig:BandsSupp}(g). 
To enable a fair comparison between the flat and dispersive band cases, Fig.~\ref{fig:ReturnProbSimSupp} presents numerically simulated return probabilities for the stroboscopic noise $\sigma=0.2$. 
The simulations show that for the dispersive band ($\theta_1=0.45\pi$, $\theta_2=0.0\pi$, $\varphi=0.2\pi$), the edge state decay is predominantly exponential (green dots), whereas for the flat band ($\theta_1=0.5\pi$, $\theta_2=0.0\pi$, $\varphi=0.2\pi$), the localization of bulk states leads to a much slower, polynomial decay at long time (red dots).
The slowing down of the decay for the dispersive band starting at about $M=100$ and its flattening at long times is an artifact due to the finite size of the lattice.
These numerical results closely matches the calculated return probabilities (black solid lines) obtained from the master equation in the finite site lattice.
These numerical findings are consistent with those reported in Ref.~\cite{rieder_localization_2018}.

\bibliography{Bibliography_new}

\begin{thebibliography}{50}%
\makeatletter
\providecommand \@ifxundefined [1]{%
 \@ifx{#1\undefined}
}%
\providecommand \@ifnum [1]{%
 \ifnum #1\expandafter \@firstoftwo
 \else \expandafter \@secondoftwo
 \fi
}%
\providecommand \@ifx [1]{%
 \ifx #1\expandafter \@firstoftwo
 \else \expandafter \@secondoftwo
 \fi
}%
\providecommand \natexlab [1]{#1}%
\providecommand \enquote  [1]{``#1''}%
\providecommand \bibnamefont  [1]{#1}%
\providecommand \bibfnamefont [1]{#1}%
\providecommand \citenamefont [1]{#1}%
\providecommand \href@noop [0]{\@secondoftwo}%
\providecommand \href [0]{\begingroup \@sanitize@url \@href}%
\providecommand \@href[1]{\@@startlink{#1}\@@href}%
\providecommand \@@href[1]{\endgroup#1\@@endlink}%
\providecommand \@sanitize@url [0]{\catcode `\\12\catcode `\$12\catcode `\&12\catcode `\#12\catcode `\^12\catcode `\_12\catcode `\%12\relax}%
\providecommand \@@startlink[1]{}%
\providecommand \@@endlink[0]{}%
\providecommand \url  [0]{\begingroup\@sanitize@url \@url }%
\providecommand \@url [1]{\endgroup\@href {#1}{\urlprefix }}%
\providecommand \urlprefix  [0]{URL }%
\providecommand \Eprint [0]{\href }%
\providecommand \doibase [0]{https://doi.org/}%
\providecommand \selectlanguage [0]{\@gobble}%
\providecommand \bibinfo  [0]{\@secondoftwo}%
\providecommand \bibfield  [0]{\@secondoftwo}%
\providecommand \translation [1]{[#1]}%
\providecommand \BibitemOpen [0]{}%
\providecommand \bibitemStop [0]{}%
\providecommand \bibitemNoStop [0]{.\EOS\space}%
\providecommand \EOS [0]{\spacefactor3000\relax}%
\providecommand \BibitemShut  [1]{\csname bibitem#1\endcsname}%
\let\auto@bib@innerbib\@empty
\bibitem [{\citenamefont {Aharonov}\ \emph {et~al.}(1993)\citenamefont {Aharonov}, \citenamefont {Davidovich},\ and\ \citenamefont {Zagury}}]{aharonov_quantum_1993}%
  \BibitemOpen
  \bibfield  {author} {\bibinfo {author} {\bibfnamefont {Y.}~\bibnamefont {Aharonov}}, \bibinfo {author} {\bibfnamefont {L.}~\bibnamefont {Davidovich}},\ and\ \bibinfo {author} {\bibfnamefont {N.}~\bibnamefont {Zagury}},\ }\bibfield  {title} {\bibinfo {title} {Quantum random walks},\ }\href {https://doi.org/10.1103/PhysRevA.48.1687} {\bibfield  {journal} {\bibinfo  {journal} {Phys. Rev. A}\ }\textbf {\bibinfo {volume} {48}},\ \bibinfo {pages} {1687} (\bibinfo {year} {1993})}\BibitemShut {NoStop}%
\bibitem [{\citenamefont {Mülken}\ and\ \citenamefont {Blumen}(2011)}]{Mulken2011}%
  \BibitemOpen
  \bibfield  {author} {\bibinfo {author} {\bibfnamefont {O.}~\bibnamefont {Mülken}}\ and\ \bibinfo {author} {\bibfnamefont {A.}~\bibnamefont {Blumen}},\ }\bibfield  {title} {\bibinfo {title} {Continuous-time quantum walks: Models for coherent transport on complex networks},\ }\href {https://doi.org/https://doi.org/10.1016/j.physrep.2011.01.002} {\bibfield  {journal} {\bibinfo  {journal} {Physics Reports}\ }\textbf {\bibinfo {volume} {502}},\ \bibinfo {pages} {37} (\bibinfo {year} {2011})}\BibitemShut {NoStop}%
\bibitem [{\citenamefont {Schreiber}\ \emph {et~al.}(2011)\citenamefont {Schreiber}, \citenamefont {Cassemiro}, \citenamefont {Poto\ifmmode~\check{c}\else \v{c}\fi{}ek}, \citenamefont {G\'abris}, \citenamefont {Jex},\ and\ \citenamefont {Silberhorn}}]{Schreiber2011}%
  \BibitemOpen
  \bibfield  {author} {\bibinfo {author} {\bibfnamefont {A.}~\bibnamefont {Schreiber}}, \bibinfo {author} {\bibfnamefont {K.~N.}\ \bibnamefont {Cassemiro}}, \bibinfo {author} {\bibfnamefont {V.}~\bibnamefont {Poto\ifmmode~\check{c}\else \v{c}\fi{}ek}}, \bibinfo {author} {\bibfnamefont {A.}~\bibnamefont {G\'abris}}, \bibinfo {author} {\bibfnamefont {I.}~\bibnamefont {Jex}},\ and\ \bibinfo {author} {\bibfnamefont {C.}~\bibnamefont {Silberhorn}},\ }\bibfield  {title} {\bibinfo {title} {Decoherence and disorder in quantum walks: From ballistic spread to localization},\ }\href {https://doi.org/10.1103/PhysRevLett.106.180403} {\bibfield  {journal} {\bibinfo  {journal} {Phys. Rev. Lett.}\ }\textbf {\bibinfo {volume} {106}},\ \bibinfo {pages} {180403} (\bibinfo {year} {2011})}\BibitemShut {NoStop}%
\bibitem [{\citenamefont {Ambainis}(2003)}]{Ambainis2003}%
  \BibitemOpen
  \bibfield  {author} {\bibinfo {author} {\bibfnamefont {A.}~\bibnamefont {Ambainis}},\ }\bibfield  {title} {\bibinfo {title} {Quantum walks and their algorithmic applications},\ }\href {https://doi.org/10.1142/S0219749903000383} {\bibfield  {journal} {\bibinfo  {journal} {International Journal of Quantum Information}\ }\textbf {\bibinfo {volume} {01}},\ \bibinfo {pages} {507} (\bibinfo {year} {2003})}\BibitemShut {NoStop}%
\bibitem [{\citenamefont {Shenvi}\ \emph {et~al.}(2003)\citenamefont {Shenvi}, \citenamefont {Kempe},\ and\ \citenamefont {Whaley}}]{shenvi_quantum_2003}%
  \BibitemOpen
  \bibfield  {author} {\bibinfo {author} {\bibfnamefont {N.}~\bibnamefont {Shenvi}}, \bibinfo {author} {\bibfnamefont {J.}~\bibnamefont {Kempe}},\ and\ \bibinfo {author} {\bibfnamefont {K.~B.}\ \bibnamefont {Whaley}},\ }\bibfield  {title} {\bibinfo {title} {Quantum random-walk search algorithm},\ }\href {https://doi.org/10.1103/PhysRevA.67.052307} {\bibfield  {journal} {\bibinfo  {journal} {Phys. Rev. A}\ }\textbf {\bibinfo {volume} {67}},\ \bibinfo {pages} {052307} (\bibinfo {year} {2003})}\BibitemShut {NoStop}%
\bibitem [{\citenamefont {Childs}(2009)}]{childs_universal_2009}%
  \BibitemOpen
  \bibfield  {author} {\bibinfo {author} {\bibfnamefont {A.~M.}\ \bibnamefont {Childs}},\ }\bibfield  {title} {\bibinfo {title} {Universal {Computation} by {Quantum} {Walk}},\ }\href {https://doi.org/10.1103/PhysRevLett.102.180501} {\bibfield  {journal} {\bibinfo  {journal} {Phys. Rev. Lett.}\ }\textbf {\bibinfo {volume} {102}},\ \bibinfo {pages} {180501} (\bibinfo {year} {2009})}\BibitemShut {NoStop}%
\bibitem [{\citenamefont {Prokof'ev}\ and\ \citenamefont {Stamp}(2006)}]{Stamp_QWalks_Decoherence_2006}%
  \BibitemOpen
  \bibfield  {author} {\bibinfo {author} {\bibfnamefont {N.~V.}\ \bibnamefont {Prokof'ev}}\ and\ \bibinfo {author} {\bibfnamefont {P.~C.~E.}\ \bibnamefont {Stamp}},\ }\bibfield  {title} {\bibinfo {title} {Decoherence and quantum walks: Anomalous diffusion and ballistic tails},\ }\href {https://doi.org/10.1103/PhysRevA.74.020102} {\bibfield  {journal} {\bibinfo  {journal} {Phys. Rev. A}\ }\textbf {\bibinfo {volume} {74}},\ \bibinfo {pages} {020102} (\bibinfo {year} {2006})}\BibitemShut {NoStop}%
\bibitem [{\citenamefont {Hines}\ and\ \citenamefont {Stamp}(2007)}]{Stamp_QWalks_Encoding_2007}%
  \BibitemOpen
  \bibfield  {author} {\bibinfo {author} {\bibfnamefont {A.~P.}\ \bibnamefont {Hines}}\ and\ \bibinfo {author} {\bibfnamefont {P.~C.~E.}\ \bibnamefont {Stamp}},\ }\bibfield  {title} {\bibinfo {title} {Quantum walks, quantum gates, and quantum computers},\ }\href {https://doi.org/10.1103/PhysRevA.75.062321} {\bibfield  {journal} {\bibinfo  {journal} {Phys. Rev. A}\ }\textbf {\bibinfo {volume} {75}},\ \bibinfo {pages} {062321} (\bibinfo {year} {2007})}\BibitemShut {NoStop}%
\bibitem [{\citenamefont {Schreiber}\ \emph {et~al.}(2012)\citenamefont {Schreiber}, \citenamefont {Gábris}, \citenamefont {Rohde}, \citenamefont {Laiho}, \citenamefont {Štefaňák}, \citenamefont {Potoček}, \citenamefont {Hamilton}, \citenamefont {Jex},\ and\ \citenamefont {Silberhorn}}]{Schreiber_2D_QW_2012}%
  \BibitemOpen
  \bibfield  {author} {\bibinfo {author} {\bibfnamefont {A.}~\bibnamefont {Schreiber}}, \bibinfo {author} {\bibfnamefont {A.}~\bibnamefont {Gábris}}, \bibinfo {author} {\bibfnamefont {P.~P.}\ \bibnamefont {Rohde}}, \bibinfo {author} {\bibfnamefont {K.}~\bibnamefont {Laiho}}, \bibinfo {author} {\bibfnamefont {M.}~\bibnamefont {Štefaňák}}, \bibinfo {author} {\bibfnamefont {V.}~\bibnamefont {Potoček}}, \bibinfo {author} {\bibfnamefont {C.}~\bibnamefont {Hamilton}}, \bibinfo {author} {\bibfnamefont {I.}~\bibnamefont {Jex}},\ and\ \bibinfo {author} {\bibfnamefont {C.}~\bibnamefont {Silberhorn}},\ }\bibfield  {title} {\bibinfo {title} {A 2d quantum walk simulation of two-particle dynamics},\ }\href {https://doi.org/10.1126/science.1218448} {\bibfield  {journal} {\bibinfo  {journal} {Science}\ }\textbf {\bibinfo {volume} {336}},\ \bibinfo {pages} {55} (\bibinfo {year} {2012})}\BibitemShut {NoStop}%
\bibitem [{\citenamefont {Lee}\ \emph {et~al.}(2015)\citenamefont {Lee}, \citenamefont {Kurzy\ifmmode~\acute{n}\else \'{n}\fi{}ski},\ and\ \citenamefont {Nha}}]{lee_quantum_2015}%
  \BibitemOpen
  \bibfield  {author} {\bibinfo {author} {\bibfnamefont {C.-W.}\ \bibnamefont {Lee}}, \bibinfo {author} {\bibfnamefont {P.}~\bibnamefont {Kurzy\ifmmode~\acute{n}\else \'{n}\fi{}ski}},\ and\ \bibinfo {author} {\bibfnamefont {H.}~\bibnamefont {Nha}},\ }\bibfield  {title} {\bibinfo {title} {Quantum walk as a simulator of nonlinear dynamics: Nonlinear dirac equation and solitons},\ }\href {https://doi.org/10.1103/PhysRevA.92.052336} {\bibfield  {journal} {\bibinfo  {journal} {Phys. Rev. A}\ }\textbf {\bibinfo {volume} {92}},\ \bibinfo {pages} {052336} (\bibinfo {year} {2015})}\BibitemShut {NoStop}%
\bibitem [{\citenamefont {Vakulchyk}\ \emph {et~al.}(2019)\citenamefont {Vakulchyk}, \citenamefont {Fistul},\ and\ \citenamefont {Flach}}]{vakulchyk_wave_2019}%
  \BibitemOpen
  \bibfield  {author} {\bibinfo {author} {\bibfnamefont {I.}~\bibnamefont {Vakulchyk}}, \bibinfo {author} {\bibfnamefont {M.~V.}\ \bibnamefont {Fistul}},\ and\ \bibinfo {author} {\bibfnamefont {S.}~\bibnamefont {Flach}},\ }\bibfield  {title} {\bibinfo {title} {Wave {Packet} {Spreading} with {Disordered} {Nonlinear} {Discrete}-{Time} {Quantum} {Walks}},\ }\href {https://doi.org/10.1103/PhysRevLett.122.040501} {\bibfield  {journal} {\bibinfo  {journal} {Phys. Rev. Lett.}\ }\textbf {\bibinfo {volume} {122}},\ \bibinfo {pages} {040501} (\bibinfo {year} {2019})}\BibitemShut {NoStop}%
\bibitem [{\citenamefont {Kok}\ \emph {et~al.}(2007)\citenamefont {Kok}, \citenamefont {Munro}, \citenamefont {Nemoto}, \citenamefont {Ralph}, \citenamefont {Dowling},\ and\ \citenamefont {Milburn}}]{kok_linear_2007}%
  \BibitemOpen
  \bibfield  {author} {\bibinfo {author} {\bibfnamefont {P.}~\bibnamefont {Kok}}, \bibinfo {author} {\bibfnamefont {W.~J.}\ \bibnamefont {Munro}}, \bibinfo {author} {\bibfnamefont {K.}~\bibnamefont {Nemoto}}, \bibinfo {author} {\bibfnamefont {T.~C.}\ \bibnamefont {Ralph}}, \bibinfo {author} {\bibfnamefont {J.~P.}\ \bibnamefont {Dowling}},\ and\ \bibinfo {author} {\bibfnamefont {G.~J.}\ \bibnamefont {Milburn}},\ }\bibfield  {title} {\bibinfo {title} {Linear optical quantum computing with photonic qubits},\ }\href {https://doi.org/10.1103/RevModPhys.79.135} {\bibfield  {journal} {\bibinfo  {journal} {Rev. Mod. Phys.}\ }\textbf {\bibinfo {volume} {79}},\ \bibinfo {pages} {135} (\bibinfo {year} {2007})}\BibitemShut {NoStop}%
\bibitem [{\citenamefont {Nahum}\ \emph {et~al.}(2018)\citenamefont {Nahum}, \citenamefont {Vijay},\ and\ \citenamefont {Haah}}]{nahum_operator_2018}%
  \BibitemOpen
  \bibfield  {author} {\bibinfo {author} {\bibfnamefont {A.}~\bibnamefont {Nahum}}, \bibinfo {author} {\bibfnamefont {S.}~\bibnamefont {Vijay}},\ and\ \bibinfo {author} {\bibfnamefont {J.}~\bibnamefont {Haah}},\ }\bibfield  {title} {\bibinfo {title} {Operator {Spreading} in {Random} {Unitary} {Circuits}},\ }\href {https://doi.org/10.1103/PhysRevX.8.021014} {\bibfield  {journal} {\bibinfo  {journal} {Phys. Rev. X}\ }\textbf {\bibinfo {volume} {8}},\ \bibinfo {pages} {021014} (\bibinfo {year} {2018})}\BibitemShut {NoStop}%
\bibitem [{\citenamefont {Monika}\ \emph {et~al.}(2025)\citenamefont {Monika}, \citenamefont {Nosrati}, \citenamefont {George}, \citenamefont {Sciara}, \citenamefont {Fazili}, \citenamefont {Marques~Muniz}, \citenamefont {Bisianov}, \citenamefont {Lo~Franco}, \citenamefont {Munro}, \citenamefont {Chemnitz}, \citenamefont {Peschel},\ and\ \citenamefont {Morandotti}}]{monika_quantum_2025}%
  \BibitemOpen
  \bibfield  {author} {\bibinfo {author} {\bibfnamefont {M.}~\bibnamefont {Monika}}, \bibinfo {author} {\bibfnamefont {F.}~\bibnamefont {Nosrati}}, \bibinfo {author} {\bibfnamefont {A.}~\bibnamefont {George}}, \bibinfo {author} {\bibfnamefont {S.}~\bibnamefont {Sciara}}, \bibinfo {author} {\bibfnamefont {R.}~\bibnamefont {Fazili}}, \bibinfo {author} {\bibfnamefont {A.~L.}\ \bibnamefont {Marques~Muniz}}, \bibinfo {author} {\bibfnamefont {A.}~\bibnamefont {Bisianov}}, \bibinfo {author} {\bibfnamefont {R.}~\bibnamefont {Lo~Franco}}, \bibinfo {author} {\bibfnamefont {W.~J.}\ \bibnamefont {Munro}}, \bibinfo {author} {\bibfnamefont {M.}~\bibnamefont {Chemnitz}}, \bibinfo {author} {\bibfnamefont {U.}~\bibnamefont {Peschel}},\ and\ \bibinfo {author} {\bibfnamefont {R.}~\bibnamefont {Morandotti}},\ }\bibfield  {title} {\bibinfo {title} {Quantum state processing through controllable synthetic temporal photonic lattices},\ }\href {https://doi.org/10.1038/s41566-024-01546-4} {\bibfield  {journal} {\bibinfo  {journal}
  {Nat. Photon.}\ }\textbf {\bibinfo {volume} {19}},\ \bibinfo {pages} {95} (\bibinfo {year} {2025})}\BibitemShut {NoStop}%
\bibitem [{\citenamefont {Schreiber}\ \emph {et~al.}(2010)\citenamefont {Schreiber}, \citenamefont {Cassemiro}, \citenamefont {Potoček}, \citenamefont {Gábris}, \citenamefont {Mosley}, \citenamefont {Andersson}, \citenamefont {Jex},\ and\ \citenamefont {Silberhorn}}]{schreiber_photons_2010}%
  \BibitemOpen
  \bibfield  {author} {\bibinfo {author} {\bibfnamefont {A.}~\bibnamefont {Schreiber}}, \bibinfo {author} {\bibfnamefont {K.~N.}\ \bibnamefont {Cassemiro}}, \bibinfo {author} {\bibfnamefont {V.}~\bibnamefont {Potoček}}, \bibinfo {author} {\bibfnamefont {A.}~\bibnamefont {Gábris}}, \bibinfo {author} {\bibfnamefont {P.~J.}\ \bibnamefont {Mosley}}, \bibinfo {author} {\bibfnamefont {E.}~\bibnamefont {Andersson}}, \bibinfo {author} {\bibfnamefont {I.}~\bibnamefont {Jex}},\ and\ \bibinfo {author} {\bibfnamefont {C.}~\bibnamefont {Silberhorn}},\ }\bibfield  {title} {\bibinfo {title} {Photons {Walking} the {Line}: {A} {Quantum} {Walk} with {Adjustable} {Coin} {Operations}},\ }\href {https://doi.org/10.1103/PhysRevLett.104.050502} {\bibfield  {journal} {\bibinfo  {journal} {Phys. Rev. Lett.}\ }\textbf {\bibinfo {volume} {104}},\ \bibinfo {pages} {050502} (\bibinfo {year} {2010})}\BibitemShut {NoStop}%
\bibitem [{\citenamefont {Wimmer}\ \emph {et~al.}(2015)\citenamefont {Wimmer}, \citenamefont {Miri}, \citenamefont {Christodoulides},\ and\ \citenamefont {Peschel}}]{Wimmer2015}%
  \BibitemOpen
  \bibfield  {author} {\bibinfo {author} {\bibfnamefont {M.}~\bibnamefont {Wimmer}}, \bibinfo {author} {\bibfnamefont {M.-A.}\ \bibnamefont {Miri}}, \bibinfo {author} {\bibfnamefont {D.}~\bibnamefont {Christodoulides}},\ and\ \bibinfo {author} {\bibfnamefont {U.}~\bibnamefont {Peschel}},\ }\bibfield  {title} {\bibinfo {title} {Observation of {Bloch} oscillations in complex {PT}-symmetric photonic lattices},\ }\href {http://dx.doi.org/10.1038/srep17760} {\bibfield  {journal} {\bibinfo  {journal} {Scientific Reports}\ }\textbf {\bibinfo {volume} {5}},\ \bibinfo {pages} {17760} (\bibinfo {year} {2015})}\BibitemShut {NoStop}%
\bibitem [{\citenamefont {Adiyatullin}\ \emph {et~al.}(2022)\citenamefont {Adiyatullin}, \citenamefont {Upreti}, \citenamefont {Lechevalier}, \citenamefont {Evain}, \citenamefont {Copie}, \citenamefont {Suret}, \citenamefont {Randoux}, \citenamefont {Delplace},\ and\ \citenamefont {Amo}}]{adiyatullin_topological_2022}%
  \BibitemOpen
  \bibfield  {author} {\bibinfo {author} {\bibfnamefont {A.~F.}\ \bibnamefont {Adiyatullin}}, \bibinfo {author} {\bibfnamefont {L.~K.}\ \bibnamefont {Upreti}}, \bibinfo {author} {\bibfnamefont {C.}~\bibnamefont {Lechevalier}}, \bibinfo {author} {\bibfnamefont {C.}~\bibnamefont {Evain}}, \bibinfo {author} {\bibfnamefont {F.}~\bibnamefont {Copie}}, \bibinfo {author} {\bibfnamefont {P.}~\bibnamefont {Suret}}, \bibinfo {author} {\bibfnamefont {S.}~\bibnamefont {Randoux}}, \bibinfo {author} {\bibfnamefont {P.}~\bibnamefont {Delplace}},\ and\ \bibinfo {author} {\bibfnamefont {A.}~\bibnamefont {Amo}},\ }\bibfield  {title} {\bibinfo {title} {Topological {Properties} of {Floquet} {Winding} {Bands} in a {Photonic} {Lattice}},\ }\href {https://doi.org/10.1103/PhysRevLett.130.056901} {\bibfield  {journal} {\bibinfo  {journal} {Phys. Rev. Lett.}\ }\textbf {\bibinfo {volume} {130}},\ \bibinfo {pages} {056901} (\bibinfo {year} {2022})}\BibitemShut {NoStop}%
\bibitem [{\citenamefont {Hu}\ \emph {et~al.}(2024)\citenamefont {Hu}, \citenamefont {Wang}, \citenamefont {Qin}, \citenamefont {Liu}, \citenamefont {Zhao}, \citenamefont {Li}, \citenamefont {Ye}, \citenamefont {Liu}, \citenamefont {Longhi}, \citenamefont {Lu},\ and\ \citenamefont {Wang}}]{hu_observing_2024}%
  \BibitemOpen
  \bibfield  {author} {\bibinfo {author} {\bibfnamefont {X.}~\bibnamefont {Hu}}, \bibinfo {author} {\bibfnamefont {S.}~\bibnamefont {Wang}}, \bibinfo {author} {\bibfnamefont {C.}~\bibnamefont {Qin}}, \bibinfo {author} {\bibfnamefont {C.}~\bibnamefont {Liu}}, \bibinfo {author} {\bibfnamefont {L.}~\bibnamefont {Zhao}}, \bibinfo {author} {\bibfnamefont {Y.}~\bibnamefont {Li}}, \bibinfo {author} {\bibfnamefont {H.}~\bibnamefont {Ye}}, \bibinfo {author} {\bibfnamefont {W.}~\bibnamefont {Liu}}, \bibinfo {author} {\bibfnamefont {S.}~\bibnamefont {Longhi}}, \bibinfo {author} {\bibfnamefont {P.}~\bibnamefont {Lu}},\ and\ \bibinfo {author} {\bibfnamefont {B.}~\bibnamefont {Wang}},\ }\bibfield  {title} {\bibinfo {title} {Observing the collapse of super-{Bloch} oscillations in strong-driving photonic temporal lattices},\ }\bibfield  {journal} {\bibinfo  {journal} {Adv. Photon.}\ }\textbf {\bibinfo {volume} {6}},\ \href {https://doi.org/10.1117/1.AP.6.4.046001} {10.1117/1.AP.6.4.046001} (\bibinfo {year}
  {2024})\BibitemShut {NoStop}%
\bibitem [{\citenamefont {Zhao}\ \emph {et~al.}(2024)\citenamefont {Zhao}, \citenamefont {Wang}, \citenamefont {Qin}, \citenamefont {Liu}, \citenamefont {Liu}, \citenamefont {Hu}, \citenamefont {Li}, \citenamefont {Wang},\ and\ \citenamefont {Lu}}]{zhao_blochzener_2024}%
  \BibitemOpen
  \bibfield  {author} {\bibinfo {author} {\bibfnamefont {L.}~\bibnamefont {Zhao}}, \bibinfo {author} {\bibfnamefont {S.}~\bibnamefont {Wang}}, \bibinfo {author} {\bibfnamefont {C.}~\bibnamefont {Qin}}, \bibinfo {author} {\bibfnamefont {Z.}~\bibnamefont {Liu}}, \bibinfo {author} {\bibfnamefont {C.}~\bibnamefont {Liu}}, \bibinfo {author} {\bibfnamefont {X.}~\bibnamefont {Hu}}, \bibinfo {author} {\bibfnamefont {Y.}~\bibnamefont {Li}}, \bibinfo {author} {\bibfnamefont {B.}~\bibnamefont {Wang}},\ and\ \bibinfo {author} {\bibfnamefont {P.}~\bibnamefont {Lu}},\ }\bibfield  {title} {\bibinfo {title} {Bloch–{Zener} oscillation with engineered {Floquet} energy bands in synthetic temporal lattices},\ }\href {https://doi.org/10.1364/OL.543457} {\bibfield  {journal} {\bibinfo  {journal} {Opt. Lett.}\ }\textbf {\bibinfo {volume} {49}},\ \bibinfo {pages} {7028} (\bibinfo {year} {2024})}\BibitemShut {NoStop}%
\bibitem [{\citenamefont {Chalabi}\ \emph {et~al.}(2019)\citenamefont {Chalabi}, \citenamefont {Barik}, \citenamefont {Mittal}, \citenamefont {Murphy}, \citenamefont {Hafezi},\ and\ \citenamefont {Waks}}]{Chalabi2019}%
  \BibitemOpen
  \bibfield  {author} {\bibinfo {author} {\bibfnamefont {H.}~\bibnamefont {Chalabi}}, \bibinfo {author} {\bibfnamefont {S.}~\bibnamefont {Barik}}, \bibinfo {author} {\bibfnamefont {S.}~\bibnamefont {Mittal}}, \bibinfo {author} {\bibfnamefont {T.~E.}\ \bibnamefont {Murphy}}, \bibinfo {author} {\bibfnamefont {M.}~\bibnamefont {Hafezi}},\ and\ \bibinfo {author} {\bibfnamefont {E.}~\bibnamefont {Waks}},\ }\bibfield  {title} {\bibinfo {title} {Synthetic {Gauge} {Field} for {Two}-{Dimensional} {Time}-{Multiplexed} {Quantum} {Random} {Walks}},\ }\href {https://doi.org/10.1103/PhysRevLett.123.150503} {\bibfield  {journal} {\bibinfo  {journal} {Physical Review Letters}\ }\textbf {\bibinfo {volume} {123}},\ \bibinfo {pages} {150503} (\bibinfo {year} {2019})}\BibitemShut {NoStop}%
\bibitem [{\citenamefont {Ye}\ \emph {et~al.}(2023)\citenamefont {Ye}, \citenamefont {Qin}, \citenamefont {Wang}, \citenamefont {Zhao}, \citenamefont {Liu}, \citenamefont {Wang}, \citenamefont {Longhi},\ and\ \citenamefont {Lu}}]{ye_reconfigurable_2023}%
  \BibitemOpen
  \bibfield  {author} {\bibinfo {author} {\bibfnamefont {H.}~\bibnamefont {Ye}}, \bibinfo {author} {\bibfnamefont {C.}~\bibnamefont {Qin}}, \bibinfo {author} {\bibfnamefont {S.}~\bibnamefont {Wang}}, \bibinfo {author} {\bibfnamefont {L.}~\bibnamefont {Zhao}}, \bibinfo {author} {\bibfnamefont {W.}~\bibnamefont {Liu}}, \bibinfo {author} {\bibfnamefont {B.}~\bibnamefont {Wang}}, \bibinfo {author} {\bibfnamefont {S.}~\bibnamefont {Longhi}},\ and\ \bibinfo {author} {\bibfnamefont {P.}~\bibnamefont {Lu}},\ }\bibfield  {title} {\bibinfo {title} {Reconfigurable refraction manipulation at synthetic temporal interfaces with scalar and vector gauge potentials},\ }\href {https://doi.org/10.1073/pnas.2300860120} {\bibfield  {journal} {\bibinfo  {journal} {Proc. Natl. Acad. Sci. U.S.A.}\ }\textbf {\bibinfo {volume} {120}},\ \bibinfo {pages} {e2300860120} (\bibinfo {year} {2023})}\BibitemShut {NoStop}%
\bibitem [{\citenamefont {Lin}\ \emph {et~al.}(2023)\citenamefont {Lin}, \citenamefont {Yi},\ and\ \citenamefont {Xue}}]{lin_manipulating_2023}%
  \BibitemOpen
  \bibfield  {author} {\bibinfo {author} {\bibfnamefont {Q.}~\bibnamefont {Lin}}, \bibinfo {author} {\bibfnamefont {W.}~\bibnamefont {Yi}},\ and\ \bibinfo {author} {\bibfnamefont {P.}~\bibnamefont {Xue}},\ }\bibfield  {title} {\bibinfo {title} {Manipulating directional flow in a two-dimensional photonic quantum walk under a synthetic magnetic field},\ }\href {https://doi.org/10.1038/s41467-023-42045-4} {\bibfield  {journal} {\bibinfo  {journal} {Nat. Commun.}\ }\textbf {\bibinfo {volume} {14}},\ \bibinfo {pages} {6283} (\bibinfo {year} {2023})}\BibitemShut {NoStop}%
\bibitem [{\citenamefont {Kitagawa}\ \emph {et~al.}(2012)\citenamefont {Kitagawa}, \citenamefont {Broome}, \citenamefont {Fedrizzi}, \citenamefont {Rudner}, \citenamefont {Berg}, \citenamefont {Kassal}, \citenamefont {Aspuru-Guzik}, \citenamefont {Demler},\ and\ \citenamefont {White}}]{kitagawa_observation_2012}%
  \BibitemOpen
  \bibfield  {author} {\bibinfo {author} {\bibfnamefont {T.}~\bibnamefont {Kitagawa}}, \bibinfo {author} {\bibfnamefont {M.~A.}\ \bibnamefont {Broome}}, \bibinfo {author} {\bibfnamefont {A.}~\bibnamefont {Fedrizzi}}, \bibinfo {author} {\bibfnamefont {M.~S.}\ \bibnamefont {Rudner}}, \bibinfo {author} {\bibfnamefont {E.}~\bibnamefont {Berg}}, \bibinfo {author} {\bibfnamefont {I.}~\bibnamefont {Kassal}}, \bibinfo {author} {\bibfnamefont {A.}~\bibnamefont {Aspuru-Guzik}}, \bibinfo {author} {\bibfnamefont {E.}~\bibnamefont {Demler}},\ and\ \bibinfo {author} {\bibfnamefont {A.~G.}\ \bibnamefont {White}},\ }\bibfield  {title} {\bibinfo {title} {Observation of topologically protected bound states in photonic quantum walks},\ }\href {https://doi.org/10.1038/ncomms1872} {\bibfield  {journal} {\bibinfo  {journal} {Nat Commun}\ }\textbf {\bibinfo {volume} {3}},\ \bibinfo {pages} {882} (\bibinfo {year} {2012})}\BibitemShut {NoStop}%
\bibitem [{\citenamefont {Bisianov}\ \emph {et~al.}(2019)\citenamefont {Bisianov}, \citenamefont {Wimmer}, \citenamefont {Peschel},\ and\ \citenamefont {Egorov}}]{Bisianov2019}%
  \BibitemOpen
  \bibfield  {author} {\bibinfo {author} {\bibfnamefont {A.}~\bibnamefont {Bisianov}}, \bibinfo {author} {\bibfnamefont {M.}~\bibnamefont {Wimmer}}, \bibinfo {author} {\bibfnamefont {U.}~\bibnamefont {Peschel}},\ and\ \bibinfo {author} {\bibfnamefont {O.~A.}\ \bibnamefont {Egorov}},\ }\bibfield  {title} {\bibinfo {title} {Stability of topologically protected edge states in nonlinear fiber loops},\ }\href {https://doi.org/10.1103/PhysRevA.100.063830} {\bibfield  {journal} {\bibinfo  {journal} {Phys. Rev. A}\ }\textbf {\bibinfo {volume} {100}},\ \bibinfo {pages} {063830} (\bibinfo {year} {2019})}\BibitemShut {NoStop}%
\bibitem [{\citenamefont {Mochizuki}\ \emph {et~al.}(2016)\citenamefont {Mochizuki}, \citenamefont {Kim},\ and\ \citenamefont {Obuse}}]{mochizuki_explicit_2016}%
  \BibitemOpen
  \bibfield  {author} {\bibinfo {author} {\bibfnamefont {K.}~\bibnamefont {Mochizuki}}, \bibinfo {author} {\bibfnamefont {D.}~\bibnamefont {Kim}},\ and\ \bibinfo {author} {\bibfnamefont {H.}~\bibnamefont {Obuse}},\ }\bibfield  {title} {\bibinfo {title} {Explicit definition of {PT} symmetry for nonunitary quantum walks with gain and loss},\ }\href {https://doi.org/10.1103/PhysRevA.93.062116} {\bibfield  {journal} {\bibinfo  {journal} {Phys. Rev. A}\ }\textbf {\bibinfo {volume} {93}},\ \bibinfo {pages} {062116} (\bibinfo {year} {2016})}\BibitemShut {NoStop}%
\bibitem [{\citenamefont {Xiao}\ \emph {et~al.}(2018)\citenamefont {Xiao}, \citenamefont {Qiu}, \citenamefont {Wang}, \citenamefont {Bian}, \citenamefont {Zhan}, \citenamefont {Obuse}, \citenamefont {Sanders}, \citenamefont {Yi},\ and\ \citenamefont {Xue}}]{xiao_higher_2018}%
  \BibitemOpen
  \bibfield  {author} {\bibinfo {author} {\bibfnamefont {L.}~\bibnamefont {Xiao}}, \bibinfo {author} {\bibfnamefont {X.}~\bibnamefont {Qiu}}, \bibinfo {author} {\bibfnamefont {K.}~\bibnamefont {Wang}}, \bibinfo {author} {\bibfnamefont {Z.}~\bibnamefont {Bian}}, \bibinfo {author} {\bibfnamefont {X.}~\bibnamefont {Zhan}}, \bibinfo {author} {\bibfnamefont {H.}~\bibnamefont {Obuse}}, \bibinfo {author} {\bibfnamefont {B.~C.}\ \bibnamefont {Sanders}}, \bibinfo {author} {\bibfnamefont {W.}~\bibnamefont {Yi}},\ and\ \bibinfo {author} {\bibfnamefont {P.}~\bibnamefont {Xue}},\ }\bibfield  {title} {\bibinfo {title} {Higher winding number in a nonunitary photonic quantum walk},\ }\href {https://doi.org/10.1103/PhysRevA.98.063847} {\bibfield  {journal} {\bibinfo  {journal} {Physical Review A}\ }\textbf {\bibinfo {volume} {98}},\ \bibinfo {pages} {63847} (\bibinfo {year} {2018})}\BibitemShut {NoStop}%
\bibitem [{\citenamefont {Xiao}\ \emph {et~al.}(2020)\citenamefont {Xiao}, \citenamefont {Deng}, \citenamefont {Wang}, \citenamefont {Zhu}, \citenamefont {Wang}, \citenamefont {Yi},\ and\ \citenamefont {Xue}}]{xiao_non-hermitian_2020}%
  \BibitemOpen
  \bibfield  {author} {\bibinfo {author} {\bibfnamefont {L.}~\bibnamefont {Xiao}}, \bibinfo {author} {\bibfnamefont {T.}~\bibnamefont {Deng}}, \bibinfo {author} {\bibfnamefont {K.}~\bibnamefont {Wang}}, \bibinfo {author} {\bibfnamefont {G.}~\bibnamefont {Zhu}}, \bibinfo {author} {\bibfnamefont {Z.}~\bibnamefont {Wang}}, \bibinfo {author} {\bibfnamefont {W.}~\bibnamefont {Yi}},\ and\ \bibinfo {author} {\bibfnamefont {P.}~\bibnamefont {Xue}},\ }\bibfield  {title} {\bibinfo {title} {Non-{Hermitian} bulk–boundary correspondence in quantum dynamics},\ }\href {https://doi.org/10.1038/s41567-020-0836-6} {\bibfield  {journal} {\bibinfo  {journal} {Nat. Phys.}\ }\textbf {\bibinfo {volume} {16}},\ \bibinfo {pages} {761} (\bibinfo {year} {2020})}\BibitemShut {NoStop}%
\bibitem [{\citenamefont {Weidemann}\ \emph {et~al.}(2020{\natexlab{a}})\citenamefont {Weidemann}, \citenamefont {Kremer}, \citenamefont {Helbig}, \citenamefont {Hofmann}, \citenamefont {Stegmaier}, \citenamefont {Greiter}, \citenamefont {Thomale},\ and\ \citenamefont {Szameit}}]{Weidemann2020}%
  \BibitemOpen
  \bibfield  {author} {\bibinfo {author} {\bibfnamefont {S.}~\bibnamefont {Weidemann}}, \bibinfo {author} {\bibfnamefont {M.}~\bibnamefont {Kremer}}, \bibinfo {author} {\bibfnamefont {T.}~\bibnamefont {Helbig}}, \bibinfo {author} {\bibfnamefont {T.}~\bibnamefont {Hofmann}}, \bibinfo {author} {\bibfnamefont {A.}~\bibnamefont {Stegmaier}}, \bibinfo {author} {\bibfnamefont {M.}~\bibnamefont {Greiter}}, \bibinfo {author} {\bibfnamefont {R.}~\bibnamefont {Thomale}},\ and\ \bibinfo {author} {\bibfnamefont {A.}~\bibnamefont {Szameit}},\ }\bibfield  {title} {\bibinfo {title} {Topological funneling of light},\ }\href {https://doi.org/10.1126/science.aaz8727} {\bibfield  {journal} {\bibinfo  {journal} {Science}\ }\textbf {\bibinfo {volume} {368}},\ \bibinfo {pages} {311} (\bibinfo {year} {2020}{\natexlab{a}})}\BibitemShut {NoStop}%
\bibitem [{\citenamefont {Mittal}\ \emph {et~al.}(2021)\citenamefont {Mittal}, \citenamefont {Raj}, \citenamefont {Dey},\ and\ \citenamefont {Goyal}}]{mittal_persistence_2021}%
  \BibitemOpen
  \bibfield  {author} {\bibinfo {author} {\bibfnamefont {V.}~\bibnamefont {Mittal}}, \bibinfo {author} {\bibfnamefont {A.}~\bibnamefont {Raj}}, \bibinfo {author} {\bibfnamefont {S.}~\bibnamefont {Dey}},\ and\ \bibinfo {author} {\bibfnamefont {S.~K.}\ \bibnamefont {Goyal}},\ }\bibfield  {title} {\bibinfo {title} {Persistence of topological phases in non-{Hermitian} quantum walks},\ }\href {https://doi.org/10.1038/s41598-021-89441-8} {\bibfield  {journal} {\bibinfo  {journal} {Sci. Rep.}\ }\textbf {\bibinfo {volume} {11}},\ \bibinfo {pages} {10262} (\bibinfo {year} {2021})}\BibitemShut {NoStop}%
\bibitem [{\citenamefont {Cardano}\ \emph {et~al.}(2017)\citenamefont {Cardano}, \citenamefont {D’Errico}, \citenamefont {Dauphin}, \citenamefont {Maffei}, \citenamefont {Piccirillo}, \citenamefont {de~Lisio}, \citenamefont {De~Filippis}, \citenamefont {Cataudella}, \citenamefont {Santamato}, \citenamefont {Marrucci}, \citenamefont {Lewenstein},\ and\ \citenamefont {Massignan}}]{Cardano2017}%
  \BibitemOpen
  \bibfield  {author} {\bibinfo {author} {\bibfnamefont {F.}~\bibnamefont {Cardano}}, \bibinfo {author} {\bibfnamefont {A.}~\bibnamefont {D’Errico}}, \bibinfo {author} {\bibfnamefont {A.}~\bibnamefont {Dauphin}}, \bibinfo {author} {\bibfnamefont {M.}~\bibnamefont {Maffei}}, \bibinfo {author} {\bibfnamefont {B.}~\bibnamefont {Piccirillo}}, \bibinfo {author} {\bibfnamefont {C.}~\bibnamefont {de~Lisio}}, \bibinfo {author} {\bibfnamefont {G.}~\bibnamefont {De~Filippis}}, \bibinfo {author} {\bibfnamefont {V.}~\bibnamefont {Cataudella}}, \bibinfo {author} {\bibfnamefont {E.}~\bibnamefont {Santamato}}, \bibinfo {author} {\bibfnamefont {L.}~\bibnamefont {Marrucci}}, \bibinfo {author} {\bibfnamefont {M.}~\bibnamefont {Lewenstein}},\ and\ \bibinfo {author} {\bibfnamefont {P.}~\bibnamefont {Massignan}},\ }\bibfield  {title} {\bibinfo {title} {Detection of {Zak} phases and topological invariants in a chiral quantum walk of twisted photons},\ }\href {https://doi.org/10.1038/ncomms15516} {\bibfield  {journal} {\bibinfo
  {journal} {Nature Communications}\ }\textbf {\bibinfo {volume} {8}},\ \bibinfo {pages} {15516} (\bibinfo {year} {2017})}\BibitemShut {NoStop}%
\bibitem [{\citenamefont {Maczewsky}\ \emph {et~al.}(2017)\citenamefont {Maczewsky}, \citenamefont {Zeuner}, \citenamefont {Nolte},\ and\ \citenamefont {Szameit}}]{Maczewsky2017}%
  \BibitemOpen
  \bibfield  {author} {\bibinfo {author} {\bibfnamefont {L.~J.}\ \bibnamefont {Maczewsky}}, \bibinfo {author} {\bibfnamefont {J.~M.}\ \bibnamefont {Zeuner}}, \bibinfo {author} {\bibfnamefont {S.}~\bibnamefont {Nolte}},\ and\ \bibinfo {author} {\bibfnamefont {A.}~\bibnamefont {Szameit}},\ }\bibfield  {title} {\bibinfo {title} {Observation of photonic anomalous {Floquet} topological insulators},\ }\href {https://doi.org/10.1038/ncomms13756} {\bibfield  {journal} {\bibinfo  {journal} {Nat. Commun.}\ }\textbf {\bibinfo {volume} {8}},\ \bibinfo {pages} {13756} (\bibinfo {year} {2017})}\BibitemShut {NoStop}%
\bibitem [{\citenamefont {Mukherjee}\ \emph {et~al.}(2017)\citenamefont {Mukherjee}, \citenamefont {Spracklen}, \citenamefont {Valiente}, \citenamefont {Andersson}, \citenamefont {Öhberg}, \citenamefont {Goldman},\ and\ \citenamefont {Thomson}}]{Mukherjee2017a}%
  \BibitemOpen
  \bibfield  {author} {\bibinfo {author} {\bibfnamefont {S.}~\bibnamefont {Mukherjee}}, \bibinfo {author} {\bibfnamefont {A.}~\bibnamefont {Spracklen}}, \bibinfo {author} {\bibfnamefont {M.}~\bibnamefont {Valiente}}, \bibinfo {author} {\bibfnamefont {E.}~\bibnamefont {Andersson}}, \bibinfo {author} {\bibfnamefont {P.}~\bibnamefont {Öhberg}}, \bibinfo {author} {\bibfnamefont {N.}~\bibnamefont {Goldman}},\ and\ \bibinfo {author} {\bibfnamefont {R.~R.}\ \bibnamefont {Thomson}},\ }\bibfield  {title} {\bibinfo {title} {Experimental observation of anomalous topological edge modes in a slowly driven photonic lattice},\ }\href {https://doi.org/10.1038/ncomms13918} {\bibfield  {journal} {\bibinfo  {journal} {Nature Communications}\ }\textbf {\bibinfo {volume} {8}},\ \bibinfo {pages} {13918} (\bibinfo {year} {2017})}\BibitemShut {NoStop}%
\bibitem [{\citenamefont {Cheng}\ \emph {et~al.}(2019)\citenamefont {Cheng}, \citenamefont {Pan}, \citenamefont {Wang}, \citenamefont {Zhang}, \citenamefont {Yu}, \citenamefont {Gover}, \citenamefont {Zhang}, \citenamefont {Li}, \citenamefont {Zhou},\ and\ \citenamefont {Zhu}}]{cheng_observation_2019}%
  \BibitemOpen
  \bibfield  {author} {\bibinfo {author} {\bibfnamefont {Q.}~\bibnamefont {Cheng}}, \bibinfo {author} {\bibfnamefont {Y.}~\bibnamefont {Pan}}, \bibinfo {author} {\bibfnamefont {H.}~\bibnamefont {Wang}}, \bibinfo {author} {\bibfnamefont {C.}~\bibnamefont {Zhang}}, \bibinfo {author} {\bibfnamefont {D.}~\bibnamefont {Yu}}, \bibinfo {author} {\bibfnamefont {A.}~\bibnamefont {Gover}}, \bibinfo {author} {\bibfnamefont {H.}~\bibnamefont {Zhang}}, \bibinfo {author} {\bibfnamefont {T.}~\bibnamefont {Li}}, \bibinfo {author} {\bibfnamefont {L.}~\bibnamefont {Zhou}},\ and\ \bibinfo {author} {\bibfnamefont {S.}~\bibnamefont {Zhu}},\ }\bibfield  {title} {\bibinfo {title} {Observation of {Anomalous} $\pi$ {Modes} in {Photonic} {Floquet} {Engineering}},\ }\href {https://doi.org/10.1103/PhysRevLett.122.173901} {\bibfield  {journal} {\bibinfo  {journal} {Phys. Rev. Lett.}\ }\textbf {\bibinfo {volume} {122}},\ \bibinfo {pages} {173901} (\bibinfo {year} {2019})}\BibitemShut {NoStop}%
\bibitem [{\citenamefont {Bessho}\ \emph {et~al.}(2022)\citenamefont {Bessho}, \citenamefont {Mochizuki}, \citenamefont {Obuse},\ and\ \citenamefont {Sato}}]{bessho_extrinsic_2022}%
  \BibitemOpen
  \bibfield  {author} {\bibinfo {author} {\bibfnamefont {T.}~\bibnamefont {Bessho}}, \bibinfo {author} {\bibfnamefont {K.}~\bibnamefont {Mochizuki}}, \bibinfo {author} {\bibfnamefont {H.}~\bibnamefont {Obuse}},\ and\ \bibinfo {author} {\bibfnamefont {M.}~\bibnamefont {Sato}},\ }\bibfield  {title} {\bibinfo {title} {Extrinsic topology of {Floquet} anomalous boundary states in quantum walks},\ }\href {https://doi.org/10.1103/PhysRevB.105.094306} {\bibfield  {journal} {\bibinfo  {journal} {Phys. Rev. B}\ }\textbf {\bibinfo {volume} {105}},\ \bibinfo {pages} {094306} (\bibinfo {year} {2022})}\BibitemShut {NoStop}%
\bibitem [{\citenamefont {El~Sokhen}\ \emph {et~al.}(2024)\citenamefont {El~Sokhen}, \citenamefont {G\'omez-Le\'on}, \citenamefont {Adiyatullin}, \citenamefont {Randoux}, \citenamefont {Delplace},\ and\ \citenamefont {Amo}}]{el_sokhen_edge-dependent_2024}%
  \BibitemOpen
  \bibfield  {author} {\bibinfo {author} {\bibfnamefont {R.}~\bibnamefont {El~Sokhen}}, \bibinfo {author} {\bibfnamefont {A.}~\bibnamefont {G\'omez-Le\'on}}, \bibinfo {author} {\bibfnamefont {A.~F.}\ \bibnamefont {Adiyatullin}}, \bibinfo {author} {\bibfnamefont {S.}~\bibnamefont {Randoux}}, \bibinfo {author} {\bibfnamefont {P.}~\bibnamefont {Delplace}},\ and\ \bibinfo {author} {\bibfnamefont {A.}~\bibnamefont {Amo}},\ }\bibfield  {title} {\bibinfo {title} {Edge-dependent anomalous topology in synthetic photonic lattices subject to discrete step walks},\ }\href {https://doi.org/10.1103/PhysRevResearch.6.023282} {\bibfield  {journal} {\bibinfo  {journal} {Phys. Rev. Res.}\ }\textbf {\bibinfo {volume} {6}},\ \bibinfo {pages} {023282} (\bibinfo {year} {2024})}\BibitemShut {NoStop}%
\bibitem [{\citenamefont {Asapanna}\ \emph {et~al.}(2025)\citenamefont {Asapanna}, \citenamefont {El~Sokhen}, \citenamefont {Adiyatullin}, \citenamefont {Hainaut}, \citenamefont {Delplace}, \citenamefont {G\'omez-Le\'on},\ and\ \citenamefont {Amo}}]{asapanna_observation_2024}%
  \BibitemOpen
  \bibfield  {author} {\bibinfo {author} {\bibfnamefont {R.}~\bibnamefont {Asapanna}}, \bibinfo {author} {\bibfnamefont {R.}~\bibnamefont {El~Sokhen}}, \bibinfo {author} {\bibfnamefont {A.~F.}\ \bibnamefont {Adiyatullin}}, \bibinfo {author} {\bibfnamefont {C.}~\bibnamefont {Hainaut}}, \bibinfo {author} {\bibfnamefont {P.}~\bibnamefont {Delplace}}, \bibinfo {author} {\bibfnamefont {A.}~\bibnamefont {G\'omez-Le\'on}},\ and\ \bibinfo {author} {\bibfnamefont {A.}~\bibnamefont {Amo}},\ }\bibfield  {title} {\bibinfo {title} {Observation of extrinsic topological phases in floquet photonic lattices},\ }\href {https://doi.org/10.1103/tqyx-qyj4} {\bibfield  {journal} {\bibinfo  {journal} {Phys. Rev. Lett.}\ }\textbf {\bibinfo {volume} {134}},\ \bibinfo {pages} {256603} (\bibinfo {year} {2025})}\BibitemShut {NoStop}%
\bibitem [{\citenamefont {Shapira}\ \emph {et~al.}(2003)\citenamefont {Shapira}, \citenamefont {Biham}, \citenamefont {Bracken},\ and\ \citenamefont {Hackett}}]{Shapira2003}%
  \BibitemOpen
  \bibfield  {author} {\bibinfo {author} {\bibfnamefont {D.}~\bibnamefont {Shapira}}, \bibinfo {author} {\bibfnamefont {O.}~\bibnamefont {Biham}}, \bibinfo {author} {\bibfnamefont {A.~J.}\ \bibnamefont {Bracken}},\ and\ \bibinfo {author} {\bibfnamefont {M.}~\bibnamefont {Hackett}},\ }\bibfield  {title} {\bibinfo {title} {One-dimensional quantum walk with unitary noise},\ }\href {https://doi.org/10.1103/PhysRevA.68.062315} {\bibfield  {journal} {\bibinfo  {journal} {Phys. Rev. A}\ }\textbf {\bibinfo {volume} {68}},\ \bibinfo {pages} {062315} (\bibinfo {year} {2003})}\BibitemShut {NoStop}%
\bibitem [{\citenamefont {Chandrashekar}\ \emph {et~al.}(2007)\citenamefont {Chandrashekar}, \citenamefont {Srikanth},\ and\ \citenamefont {Banerjee}}]{Chandrashekar2007}%
  \BibitemOpen
  \bibfield  {author} {\bibinfo {author} {\bibfnamefont {C.~M.}\ \bibnamefont {Chandrashekar}}, \bibinfo {author} {\bibfnamefont {R.}~\bibnamefont {Srikanth}},\ and\ \bibinfo {author} {\bibfnamefont {S.}~\bibnamefont {Banerjee}},\ }\bibfield  {title} {\bibinfo {title} {Symmetries and noise in quantum walk},\ }\href {https://doi.org/10.1103/PhysRevA.76.022316} {\bibfield  {journal} {\bibinfo  {journal} {Phys. Rev. A}\ }\textbf {\bibinfo {volume} {76}},\ \bibinfo {pages} {022316} (\bibinfo {year} {2007})}\BibitemShut {NoStop}%
\bibitem [{\citenamefont {Yin}\ \emph {et~al.}(2008)\citenamefont {Yin}, \citenamefont {Katsanos},\ and\ \citenamefont {Evangelou}}]{Yin2008}%
  \BibitemOpen
  \bibfield  {author} {\bibinfo {author} {\bibfnamefont {Y.}~\bibnamefont {Yin}}, \bibinfo {author} {\bibfnamefont {D.~E.}\ \bibnamefont {Katsanos}},\ and\ \bibinfo {author} {\bibfnamefont {S.~N.}\ \bibnamefont {Evangelou}},\ }\bibfield  {title} {\bibinfo {title} {Quantum walks on a random environment},\ }\href {https://doi.org/10.1103/PhysRevA.77.022302} {\bibfield  {journal} {\bibinfo  {journal} {Phys. Rev. A}\ }\textbf {\bibinfo {volume} {77}},\ \bibinfo {pages} {022302} (\bibinfo {year} {2008})}\BibitemShut {NoStop}%
\bibitem [{\citenamefont {Rieder}\ \emph {et~al.}(2018)\citenamefont {Rieder}, \citenamefont {Sieberer}, \citenamefont {Fischer},\ and\ \citenamefont {Fulga}}]{rieder_localization_2018}%
  \BibitemOpen
  \bibfield  {author} {\bibinfo {author} {\bibfnamefont {M.-T.}\ \bibnamefont {Rieder}}, \bibinfo {author} {\bibfnamefont {L.~M.}\ \bibnamefont {Sieberer}}, \bibinfo {author} {\bibfnamefont {M.~H.}\ \bibnamefont {Fischer}},\ and\ \bibinfo {author} {\bibfnamefont {I.~C.}\ \bibnamefont {Fulga}},\ }\bibfield  {title} {\bibinfo {title} {Localization {Counteracts} {Decoherence} in {Noisy} {Floquet} {Topological} {Chains}},\ }\href {https://doi.org/10.1103/PhysRevLett.120.216801} {\bibfield  {journal} {\bibinfo  {journal} {Phys. Rev. Lett.}\ }\textbf {\bibinfo {volume} {120}},\ \bibinfo {pages} {216801} (\bibinfo {year} {2018})}\BibitemShut {NoStop}%
\bibitem [{\citenamefont {Sieberer}\ \emph {et~al.}(2018)\citenamefont {Sieberer}, \citenamefont {Rieder}, \citenamefont {Fischer},\ and\ \citenamefont {Fulga}}]{sieberer_statistical_2018}%
  \BibitemOpen
  \bibfield  {author} {\bibinfo {author} {\bibfnamefont {L.~M.}\ \bibnamefont {Sieberer}}, \bibinfo {author} {\bibfnamefont {M.-T.}\ \bibnamefont {Rieder}}, \bibinfo {author} {\bibfnamefont {M.~H.}\ \bibnamefont {Fischer}},\ and\ \bibinfo {author} {\bibfnamefont {I.~C.}\ \bibnamefont {Fulga}},\ }\bibfield  {title} {\bibinfo {title} {Statistical periodicity in driven quantum systems: {General} formalism and application to noisy {Floquet} topological chains},\ }\href {https://doi.org/10.1103/PhysRevB.98.214301} {\bibfield  {journal} {\bibinfo  {journal} {Phys. Rev. B}\ }\textbf {\bibinfo {volume} {98}},\ \bibinfo {pages} {214301} (\bibinfo {year} {2018})}\BibitemShut {NoStop}%
\bibitem [{\citenamefont {Regensburger}\ \emph {et~al.}(2011)\citenamefont {Regensburger}, \citenamefont {Bersch}, \citenamefont {Hinrichs}, \citenamefont {Onishchukov}, \citenamefont {Schreiber}, \citenamefont {Silberhorn},\ and\ \citenamefont {Peschel}}]{Regensburger2011}%
  \BibitemOpen
  \bibfield  {author} {\bibinfo {author} {\bibfnamefont {A.}~\bibnamefont {Regensburger}}, \bibinfo {author} {\bibfnamefont {C.}~\bibnamefont {Bersch}}, \bibinfo {author} {\bibfnamefont {B.}~\bibnamefont {Hinrichs}}, \bibinfo {author} {\bibfnamefont {G.}~\bibnamefont {Onishchukov}}, \bibinfo {author} {\bibfnamefont {A.}~\bibnamefont {Schreiber}}, \bibinfo {author} {\bibfnamefont {C.}~\bibnamefont {Silberhorn}},\ and\ \bibinfo {author} {\bibfnamefont {U.}~\bibnamefont {Peschel}},\ }\bibfield  {title} {\bibinfo {title} {Photon {Propagation} in a {Discrete} {Fiber} {Network}: {An} {Interplay} of {Coherence} and {Losses}},\ }\href {https://doi.org/10.1103/PhysRevLett.107.233902} {\bibfield  {journal} {\bibinfo  {journal} {Phys. Rev. Lett.}\ }\textbf {\bibinfo {volume} {107}},\ \bibinfo {pages} {233902} (\bibinfo {year} {2011})}\BibitemShut {NoStop}%
\bibitem [{\citenamefont {Weidemann}\ \emph {et~al.}(2020{\natexlab{b}})\citenamefont {Weidemann}, \citenamefont {Kremer}, \citenamefont {Helbig}, \citenamefont {Hofmann}, \citenamefont {Stegmaier}, \citenamefont {Greiter}, \citenamefont {Thomale},\ and\ \citenamefont {Szameit}}]{weidemann_topological_2020}%
  \BibitemOpen
  \bibfield  {author} {\bibinfo {author} {\bibfnamefont {S.}~\bibnamefont {Weidemann}}, \bibinfo {author} {\bibfnamefont {M.}~\bibnamefont {Kremer}}, \bibinfo {author} {\bibfnamefont {T.}~\bibnamefont {Helbig}}, \bibinfo {author} {\bibfnamefont {T.}~\bibnamefont {Hofmann}}, \bibinfo {author} {\bibfnamefont {A.}~\bibnamefont {Stegmaier}}, \bibinfo {author} {\bibfnamefont {M.}~\bibnamefont {Greiter}}, \bibinfo {author} {\bibfnamefont {R.}~\bibnamefont {Thomale}},\ and\ \bibinfo {author} {\bibfnamefont {A.}~\bibnamefont {Szameit}},\ }\bibfield  {title} {\bibinfo {title} {Topological funneling of light},\ }\href {https://doi.org/10.1126/science.aaz8727} {\bibfield  {journal} {\bibinfo  {journal} {Science}\ }\textbf {\bibinfo {volume} {368}},\ \bibinfo {pages} {311} (\bibinfo {year} {2020}{\natexlab{b}})}\BibitemShut {NoStop}%
\bibitem [{\citenamefont {Dinani}\ \emph {et~al.}(2025)\citenamefont {Dinani}, \citenamefont {Pyrialakos}, \citenamefont {Berman~Bradley}, \citenamefont {Monika}, \citenamefont {Ren}, \citenamefont {Selim}, \citenamefont {Peschel}, \citenamefont {Christodoulides},\ and\ \citenamefont {Khajavikhan}}]{dinani_universal_2025}%
  \BibitemOpen
  \bibfield  {author} {\bibinfo {author} {\bibfnamefont {H.~M.}\ \bibnamefont {Dinani}}, \bibinfo {author} {\bibfnamefont {G.~G.}\ \bibnamefont {Pyrialakos}}, \bibinfo {author} {\bibfnamefont {A.~M.}\ \bibnamefont {Berman~Bradley}}, \bibinfo {author} {\bibfnamefont {M.}~\bibnamefont {Monika}}, \bibinfo {author} {\bibfnamefont {H.}~\bibnamefont {Ren}}, \bibinfo {author} {\bibfnamefont {M.~A.}\ \bibnamefont {Selim}}, \bibinfo {author} {\bibfnamefont {U.}~\bibnamefont {Peschel}}, \bibinfo {author} {\bibfnamefont {D.~N.}\ \bibnamefont {Christodoulides}},\ and\ \bibinfo {author} {\bibfnamefont {M.}~\bibnamefont {Khajavikhan}},\ }\bibfield  {title} {\bibinfo {title} {Universal routing of light via optical thermodynamics},\ }\href {https://doi.org/10.1038/s41566-025-01756-4} {\bibfield  {journal} {\bibinfo  {journal} {Nat. Photon.}\ }\textbf {\bibinfo {volume} {19}},\ \bibinfo {pages} {1116} (\bibinfo {year} {2025})}\BibitemShut {NoStop}%
\bibitem [{\citenamefont {Wang}\ \emph {et~al.}(2018)\citenamefont {Wang}, \citenamefont {Xiao}, \citenamefont {Qiu}, \citenamefont {Wang}, \citenamefont {Yi},\ and\ \citenamefont {Xue}}]{wang_detecting_2018}%
  \BibitemOpen
  \bibfield  {author} {\bibinfo {author} {\bibfnamefont {X.}~\bibnamefont {Wang}}, \bibinfo {author} {\bibfnamefont {L.}~\bibnamefont {Xiao}}, \bibinfo {author} {\bibfnamefont {X.}~\bibnamefont {Qiu}}, \bibinfo {author} {\bibfnamefont {K.}~\bibnamefont {Wang}}, \bibinfo {author} {\bibfnamefont {W.}~\bibnamefont {Yi}},\ and\ \bibinfo {author} {\bibfnamefont {P.}~\bibnamefont {Xue}},\ }\bibfield  {title} {\bibinfo {title} {Detecting topological invariants and revealing topological phase transitions in discrete-time photonic quantum walks},\ }\href {https://doi.org/10.1103/PhysRevA.98.013835} {\bibfield  {journal} {\bibinfo  {journal} {Phys. Rev. A}\ }\textbf {\bibinfo {volume} {98}},\ \bibinfo {pages} {13835} (\bibinfo {year} {2018})}\BibitemShut {NoStop}%
\bibitem [{\citenamefont {Xu}\ \emph {et~al.}(2018)\citenamefont {Xu}, \citenamefont {Wang}, \citenamefont {Pan}, \citenamefont {Sun}, \citenamefont {Xu}, \citenamefont {Chen}, \citenamefont {Tang}, \citenamefont {Gong}, \citenamefont {Han}, \citenamefont {Li},\ and\ \citenamefont {Guo}}]{xu_measuring_2018}%
  \BibitemOpen
  \bibfield  {author} {\bibinfo {author} {\bibfnamefont {X.-Y.}\ \bibnamefont {Xu}}, \bibinfo {author} {\bibfnamefont {Q.-Q.}\ \bibnamefont {Wang}}, \bibinfo {author} {\bibfnamefont {W.-W.}\ \bibnamefont {Pan}}, \bibinfo {author} {\bibfnamefont {K.}~\bibnamefont {Sun}}, \bibinfo {author} {\bibfnamefont {J.-S.}\ \bibnamefont {Xu}}, \bibinfo {author} {\bibfnamefont {G.}~\bibnamefont {Chen}}, \bibinfo {author} {\bibfnamefont {J.-S.}\ \bibnamefont {Tang}}, \bibinfo {author} {\bibfnamefont {M.}~\bibnamefont {Gong}}, \bibinfo {author} {\bibfnamefont {Y.-J.}\ \bibnamefont {Han}}, \bibinfo {author} {\bibfnamefont {C.-F.}\ \bibnamefont {Li}},\ and\ \bibinfo {author} {\bibfnamefont {G.-C.}\ \bibnamefont {Guo}},\ }\bibfield  {title} {\bibinfo {title} {Measuring the {Winding} {Number} in a {Large}-{Scale} {Chiral} {Quantum} {Walk}},\ }\href {https://doi.org/10.1103/PhysRevLett.120.260501} {\bibfield  {journal} {\bibinfo  {journal} {Phys. Rev. Lett.}\ }\textbf {\bibinfo {volume} {120}},\ \bibinfo {pages} {260501}
  (\bibinfo {year} {2018})}\BibitemShut {NoStop}%
\bibitem [{\citenamefont {Bogaerts}\ \emph {et~al.}(2020)\citenamefont {Bogaerts}, \citenamefont {Pérez}, \citenamefont {Capmany}, \citenamefont {Miller}, \citenamefont {Poon}, \citenamefont {Englund}, \citenamefont {Morichetti},\ and\ \citenamefont {Melloni}}]{bogaerts_programmable_2020}%
  \BibitemOpen
  \bibfield  {author} {\bibinfo {author} {\bibfnamefont {W.}~\bibnamefont {Bogaerts}}, \bibinfo {author} {\bibfnamefont {D.}~\bibnamefont {Pérez}}, \bibinfo {author} {\bibfnamefont {J.}~\bibnamefont {Capmany}}, \bibinfo {author} {\bibfnamefont {D.~A.~B.}\ \bibnamefont {Miller}}, \bibinfo {author} {\bibfnamefont {J.}~\bibnamefont {Poon}}, \bibinfo {author} {\bibfnamefont {D.}~\bibnamefont {Englund}}, \bibinfo {author} {\bibfnamefont {F.}~\bibnamefont {Morichetti}},\ and\ \bibinfo {author} {\bibfnamefont {A.}~\bibnamefont {Melloni}},\ }\bibfield  {title} {\bibinfo {title} {Programmable photonic circuits},\ }\href {https://doi.org/10.1038/s41586-020-2764-0} {\bibfield  {journal} {\bibinfo  {journal} {Nature}\ }\textbf {\bibinfo {volume} {586}},\ \bibinfo {pages} {207} (\bibinfo {year} {2020})}\BibitemShut {NoStop}%
\bibitem [{\citenamefont {Afzal}\ \emph {et~al.}(2020)\citenamefont {Afzal}, \citenamefont {Zimmerling}, \citenamefont {Ren}, \citenamefont {Perron},\ and\ \citenamefont {Van}}]{afzal_realization_2020}%
  \BibitemOpen
  \bibfield  {author} {\bibinfo {author} {\bibfnamefont {S.}~\bibnamefont {Afzal}}, \bibinfo {author} {\bibfnamefont {T.~J.}\ \bibnamefont {Zimmerling}}, \bibinfo {author} {\bibfnamefont {Y.}~\bibnamefont {Ren}}, \bibinfo {author} {\bibfnamefont {D.}~\bibnamefont {Perron}},\ and\ \bibinfo {author} {\bibfnamefont {V.}~\bibnamefont {Van}},\ }\bibfield  {title} {\bibinfo {title} {Realization of {Anomalous} {Floquet} {Insulators} in {Strongly} {Coupled} {Nanophotonic} {Lattices}},\ }\href {https://doi.org/10.1103/PhysRevLett.124.253601} {\bibfield  {journal} {\bibinfo  {journal} {Phys. Rev. Lett.}\ }\textbf {\bibinfo {volume} {124}},\ \bibinfo {pages} {253601} (\bibinfo {year} {2020})}\BibitemShut {NoStop}%
\bibitem [{\citenamefont {Wimmer}\ \emph {et~al.}(2017)\citenamefont {Wimmer}, \citenamefont {Price}, \citenamefont {Carusotto},\ and\ \citenamefont {Peschel}}]{Wimmer2017}%
  \BibitemOpen
  \bibfield  {author} {\bibinfo {author} {\bibfnamefont {M.}~\bibnamefont {Wimmer}}, \bibinfo {author} {\bibfnamefont {H.~M.}\ \bibnamefont {Price}}, \bibinfo {author} {\bibfnamefont {I.}~\bibnamefont {Carusotto}},\ and\ \bibinfo {author} {\bibfnamefont {U.}~\bibnamefont {Peschel}},\ }\bibfield  {title} {\bibinfo {title} {Experimental measurement of the {Berry} curvature from anomalous transport},\ }\href {http://dx.doi.org/10.1038/nphys4050} {\bibfield  {journal} {\bibinfo  {journal} {Nature Physics}\ }\textbf {\bibinfo {volume} {13}},\ \bibinfo {pages} {545} (\bibinfo {year} {2017})}\BibitemShut {NoStop}%
\bibitem [{Dat()}]{Data}%
  \BibitemOpen
  \href {https://doi.org/10.57745/7MIDRA} {\bibinfo {title} {The data sets correspoding to the figures of this article are available at doi:10.57745/7midra.}}\BibitemShut {Stop}%
\end{thebibliography}%
\end{document}